\documentclass[aps,pra,epsfig]{revtex4}

\usepackage[dvips]{graphics}
\usepackage{graphicx}
\usepackage{amsfonts}
\usepackage{amssymb}
\usepackage{amsmath}

\begin{document}

\title{Inverse Ising problem for one-dimensional chains with 
arbitrary finite-range couplings}

\author{Giacomo Gori}
\affiliation{SISSA, Via Bonomea 265, I-34136, Trieste, Italy}
\affiliation{INFN, Sezione di Trieste, I-34127 Trieste, Italy}

\author{Andrea Trombettoni}
\affiliation{SISSA, Via Bonomea 265, I-34136, Trieste, Italy}
\affiliation{INFN, Sezione di Trieste, I-34127 Trieste, Italy}

\date{October 20, 2011}

\begin{abstract}
We study Ising chains with arbitrary multispin finite-range couplings,
providing an explicit solution of the associated inverse Ising problem,
i.e. the problem of inferring the values of the coupling
constants from the correlation functions. As an application,
we reconstruct the couplings of chain Ising Hamiltonians having 
exponential or power-law two-spin plus three- or four-spin couplings.
The generalization of the method to ladders and to Ising systems 
where a mean-field interaction is added to general finite-range couplings 
is as well as discussed.
\end{abstract}
\maketitle

\section{Introduction}

Parameter estimation is a central issue in system modeling: 
a typical problem is to start from a certain amount of information 
on a given system (e.g. 
its correlation functions) and then extract the parameters of a 
model which is supposed to describe the its properties 
\cite{beck1977,bezruchko2010}. 
The parameter estimation procedure gives 
insight on the validity of the model and can suggest the introduction of more
appropriate and efficient models.

A usual approach is to extract the parameters from 
an instance of the problem in certain 
conditions and subsequently testing the model in other instances. 
From this point of view 
is useful to deal with systems in conditions where the relation between
observables and model parameters is more transparent:
e.g. for a statistical mechanics system this corresponds to high/low 
temperature or field. Once the parameters have been estimated, one moves 
to more interesting parameter regions,
where the full complexity of the system shows up. Such an approach, when
translated into the wide arena of complex systems, generally cannot be
carried out since no knob such as temperature or field is available, 
so that we may be faced with the inverse problem in the hardest region.

A huge interest in obtaining accurate parameter estimation stems from
the current availability of large datasets in several areas of biology, 
economy and social sciences,
to name a few examples DNA sequences, stock market 
time series and Internet traffic data 
(see more references in \cite{barabasi2002}).
This great amount of data has made even more pressing the quest for efficient
models, allowing us to extract and encode the relevant information.
Various techniques have been developed in order to solve this problem: two
general approaches which can be flexibly adapted to the specific problems are
Bayesian model comparison \cite{carlin1995} and Boltzmann-machine
learning \cite{mckay2003}.

In the past decade a significant contribution to
topic of parameter estimation came from the application of 
typical statistical mechanics techniques
which turned out to be very useful in the modeling and study of different
fields ranging from neurobiology \cite{schneidman2006,cocco2009,cocco2011} 
to the economy \cite{mantegna2000}.
The description of a system using statistical models 
(and in particular Ising-like models) 
appears natural in many contexts: e.g.,
effective Ising  models generally arise 
when the space of states is intrinsically discrete 
(e.g., for DNA and proteins) 
and, even when this is not the case, some
Ising variables may be lurking behind the continuous ones.
In the statistical physics realm such emergence of effective Ising models 
could occur near a critical point when the microscopic model 
is in the Ising universality class \cite{mussardo2010} - 
but one can also find more subtle examples where discrete 
Ising-like spin degrees of freedom describe some hidden order, 
e.g. the chiral ordering in frustrated continuous spin models 
\cite{villain1980,chandra1990,horiguchi1990,becca2003}.

A paradigmatic example considered by the statistical physics community
in the context of parameter estimation is, of course, the inverse Ising 
problem, i.e. the problem of inferring the values of the coupling
constants of a general Ising model from the correlation functions. 
The inverse Ising problem has been tackled by numerical and analytical 
methods, often adapting old
techniques to the problem at hand. Among these attempts we mention Monte Carlo
optimization \cite{krauth2006}, Message Passing based algorithms 
\cite{marinari2010} and Thouless-Anderson-Palmer equations
approaches \cite{roudi2009} (see \cite{sessak2010} for a review). 
Field theoretical techniques has been
used by Sessak and Monasson \cite{sessak2009} who perturbatively
calculated, in terms of the correlations, expressions for the interaction
parameters of a general (heterogeneous) Ising model 
with two-body interaction and external field.
Most of the available results on the inverse Ising problem concerns 
Ising models having two-spin interactions:
in this context exact methods, solving
the inverse Ising problem with general multispin interactions, are welcome.

We decided to concentrate in this paper on the inverse Ising problem in 
one dimension. The motivation is three-fold: firstly
one-dimensionality allows for exact solutions. In this manuscript we indeed present 
explicit analytical formulas to exactly perform the inversion for one-dimensional Ising systems 
having general multispin interactions. Our results therefore provide a theoretical laboratory where different 
approximate inverse Ising techniques \cite{krauth2006,marinari2010,roudi2009} can be benchmarked against the exact
results obtained using our method: in the following we compare some other approximate methods with exact results. 
The possibility of testing approximate methods against exact results in one-dimensional systems 
is not our only motivation: indeed one-dimensional classical models are often
employed to describe the conformational transition
of systems, as proteins or DNA, naturally possessing an
underlying one-dimensional structure. Such simple models are found to capture some of 
the global properties of these complex systems as long as conformational
properties are concerned. The existence of exact methods would then help to determine 
the parameters and the important interactions of effective models describing the properties of such systems. More in detail, 
the use of one-dimensional statistical mechanics models applied 
to systems like proteins or DNA is usually
based on the individuation of a reduced set of states representing the conformational state of a given elementary unit: 
e.g. in protein systems the states could be chosen as helix, coil and sheet 
(for aminoacids belonging to an $\alpha$-helix, to a coil and to a $\beta$-sheet, respectively). 
The task is then, given this reduced set of states, to estimate the probabilities to have
the consecutive elements in different states \cite{zimm1959,lifson1961} and then our method 
(working for Ising and Potts models) would then allow for the determination of the parameters of effective discrete models.
We notice that in our method we can consider also longer ranged couplings ({\emph i.e.}, longer than nearest-neighbour) emulating interactions among 
aminoacids distant along the chain, but near in physical space 
\cite{schreck2010}.  

Another motivation for our work is based on the fact that one-dimensional Ising-like
models can also be used to deal with stationary time-series of correlated data: as we will later discuss in the Conclusions, it is possible 
to connect stationary time-series of data by using a mapping 
onto an equilibrium discrete Markov chain having finite memory. For this application, the inversion task (to which we can refer as an 
inverse Markov problem) consists in extracting from the data the transition probabilities of the associated guessed Markov chain: 
therefore, given the similarity of the two inversion (Ising and Markov) problems, the existence of exact techniques can provide in perspective  
a way to effectively attack the inverse Markov problem. We observe that the method of using Markov chains to describe 
sequences of data may prove useful even in biological realms when statistical
properties of e.g. DNA sequences are concerned \cite{durbin2002}.

In the following we study the one-dimensional inverse Ising problem 
with general finite-range multispin interactions: by finite range $R$  
we mean that that two spins exceeding the distance $R$ do not interact 
(this implies that at maximum $R$-spin couplings can be present). 
We will then consider the reconstruction of Ising models having 
exponential or power-law two-spin couplings 
(and three- or four-spin interactions), approximating them 
with a finite range $R$ and checking the validity of the reconstructed 
couplings. A remark about dimensionality is due: 
being the dimension set to one, the system cannot 
order at finite temperature. However we show that 
mean-field like interactions can be included 
in our formalism, so that one can treat systems having finite-range 
multispin couplings and long-range mean-field interactions giving rise to 
finite-temperature transitions. Another possibility
would be to extend the range of the interaction and perform a 
so-called finite-range scaling. Such a technique has been 
employed  \cite{uzelac1988,uzelac1989} to 
the Ising model with power-law $1/r^\alpha$ decaying interactions, 
a model exhibiting a rich behaviour including a 
Berezinskii-Kosterlitz-Thouless
transition (for $\alpha=2$)
\cite{thouless1969,kosterlitz1976,cardy1981,luijten2001} and 
gaussian and non-gaussian RG fixed points (in the range of 
$\alpha$ between $1$ and $2$) as the decay exponent $\alpha$ is
varied \cite{fisher1972,bloete1997,binder2001}.

In this paper we present the solution of the inverse problem 
for a one-dimensional Ising model
with finite-range arbitrary interactions, 
i.e. not restricted to the one- and
two-body type.
The main result of our paper is formula \eqref{entropy} 
which expresses the entropy of a one-dimensional
translational invariant system (in equilibrium) in terms of a sufficiently 
large number of correlation functions, 
from which the inversion formula \eqref{inversion} 
immediately follow.

We observe that Ising chains are usually treated via
transfer matrix method, but when longer range or multispin 
types of interaction are
included the search for the parameters reproducing the observables might 
become very onerous. Our method provides a direct method 
of estimating the parameters when a sufficiently 
large number of correlation function is known.
The inclusion of many-body interactions may prove useful for the
description of complex systems where the two-body assumption is not 
justified or in more traditional many-body
systems with long-range interaction, where the construction of low-energy
effective theories quite naturally leads to the appearance of multispin
interactions \cite{micheli2007}.

The paper is structured as follows: in section 
\ref{notation_and_statement} we introduce our notations and 
we state the mathematical problem. 
Section \ref{inversion_formula} contains our main result 
on the entropy in terms of the correlation functions and the 
resulting inversion formula. The obtained result 
is illustrated on simple problems in section \ref{simple_examples}.
In section \ref{realistic_examples}
we examine more complicated examples where the usefulness of our result is
shown. We analyze models formally 
not have finite-range interactions and having 
exponential or power-law two-spin interactions plus multispin interactions.
The data generated by Monte Carlo simulations are analyzed with our technique
which correctly detects the structure of interactions. 
In section \ref{MF_LR} we briefly discuss how the developed 
formalism may be modified to allow mean-field interactions. 
Finally we draw our conclusions in
\ref{conclusions_and_perspectives}. The Appendix 
present checks of the obtained findings for small values 
of the range $R$ using the transfer matrix method, and as well as supplementary material 
on the $j_1-j_2$ Ising model.

\section{Notation and statement of the problem}\label{notation_and_statement}

We consider a general one-dimensional Ising model with multispin interactions 
defined by the Hamiltonian
\begin{displaymath}
 \mathcal H (\sigma_N)=-\sum_{i_1} j^{(1)}_{i_1} s_{i_1} -
   \sum_{(i_1,i_2)} j^{(2)}_{i_1,i_2} s_{i_1} s_{i_2}- \sum_{(i_1,i_2,i_3)} j^{(3)}_{i_1,i_2,i_3} s_{i_1} s_{i_2} s_{i_3} 
\end{displaymath}
\begin{equation}
 - \sum_{(i_1,i_2,i_3,i_4)} j^{(4)}_{i_1,i_2,i_3,i_4} s_{i_1} s_{i_2} s_{i_3} s_{i_4} - \ldots 
\label{HamiltonianExplicit}
\end{equation}
where $\sigma_N=\{s_1, s_2, \ldots, s_N\}$ is the configuration of the 
$N$ Ising spins ($s_i=\pm 1$); periodic boundary conditions 
will be assumed so that
$s_n=s_m$ for $n\equiv m \! \! \! \mod \! N$. The sums runs over distinct 
couples, triples and so on; the temperature dependence is 
absorbed in the coupling constants: explicitly, $j_{i_1}^{(1)} \equiv \beta J_{i_1}^{(1)}$, 
$j_{i_1,i_2}^{(2)} \equiv \beta J_{i_1,i_2}^{(2)}$, and so on (where {\emph e.g.} 
$J_{i_1,i_2}^{(2)}$ is the two-body coupling among a spin in $i_1$ and a spin in $i_2$ - as usual 
$\beta=1/k_B T$).

The couplings $j^{(n)}$ are assumed to 
be invariant under translation by $\rho$ spins (for simplicity we will assume 
$N/\rho$ is an integer, but since we are 
interested in the $N\rightarrow \infty$ limit this is not strictly
necessary): this condition reads 
\begin{equation}
 j^{(n)}_{i_1,i_2,\ldots,i_n}= j^{(n)}_{i_1+\rho,i_2+\rho,\ldots,i_n+\rho}
\end{equation}
(if the indices on the right hand side exceed $N$, they have to be replaced
by the indices equivalent modulo $N$ contained in the set $\{1,\ldots, N\}$).
Finally we assume that the couplings are zero 
if their indices cannot be brought by a translation
of a multiple of $\rho$ to a subset of $\{1,\ldots ,R\}$.

Since the use of the 
form \eqref{HamiltonianExplicit} of the Hamiltonian may be cumbersome, 
it is convenient introduce a more compact notation, 
rewriting Hamiltonian \eqref{HamiltonianExplicit} as
\begin{equation}
 \mathcal H (\sigma_N)=-\sum_{\mathrm{Rg}(\mu)\le R}^{'} \sum_{i=1}^{N/\rho}
j_{\mu} O_{\mu+i \rho} (\sigma_N), \label{Hamiltonian}
\end{equation}
where $\mu$ is a subset of $\{1,\ldots, R\}$ (this is encoded in the
writing $\mathrm{Rg}(\mu)\le R$, which stands for ``the range of the interaction
is less or equal than $R$''). $\rho$ is the periodicity of the interaction and  
$O_{\mu+i \rho}$ is an operator associated to the subset
$\mu=\{n_{1}, n_{2}, \ldots n_{|\mu|}\}$ ($|\mu|$ is the number of elements of
$\mu$) translated by $i \rho$ which acts on the spins as 
\begin{equation}
 O_{\mu+i \rho} (\sigma_N)=s_{n_{1} + i \rho} s_{n_{2} + i \rho} \ldots
s_{n_{|\mu|} + i \rho}.
\end{equation}
For the null subset $\varnothing$ we define $O_{\varnothing}(\cdot)=1$.
The prime in the sum over $\mu$ in \eqref{Hamiltonian} indicates that the null
subset (which would contribute just to a constant in the Hamiltonian) is not
included and that the terms related by a translation of a multiple of $\rho$ are
counted only once, in order to avoid the presence of equivalent operators in the
Hamiltonian.

Once the Hamiltonian is specified we proceed in the usual calculation of the
thermodynamic quantities, defining the partition function
\begin{equation}
 \mathcal Z_N = \sum_{\sigma_N} e^{-\mathcal H (\sigma_N)},
\end{equation}
the free energy per elementary unit cell in the infinite volume limit
(i.e. $\rho$ spins)
\begin{equation}
 f=-\lim_{N\rightarrow\infty}\frac{1}{N/\rho} \log (Z_N)
\end{equation}
and the correlation functions associated to the operator $\mu$
\begin{equation}
 g_{\mu}=\langle O_{\mu}\rangle\equiv\lim_{N\rightarrow\infty}\frac{1}{\mathcal Z}\sum_{\sigma} O_{\mu}
(\sigma_N) e^{-\mathcal H (\sigma_N)}
\label{corr}
\end{equation}
(by definition, $g_{\varnothing}=1$).

\section{Inversion formula}\label{inversion_formula}

The inverse problem for the system introduced in the previous section 
is stated as follows: given the set of
correlations $\{g_{\mu}\}$ determine the couplings $\{j_{\mu}\}$. 
The Hamiltonian is the one specified in equation \eqref{Hamiltonian},
i.e. the most general finite-range multispin Hamiltonian; 
in section \ref{MF_LR} we will extend this treatment to include 
long-range mean-field interactions.

The calculation is based on the evaluation of the entropy per unit cell
$s(\{g_{\mu}\})$ characterized by the set of correlation functions
$\{g_{\mu}\}$. Once $s(\{g_{\mu}\})$ is known we may compute the free energy
\begin{equation}
 f(\{g_{\mu}\})=e(\{g_{\mu}\})-s(\{g_{\mu}\})=-\sum_{\mathrm{Rg}(\mu)\le R}^{'}
g_{\mu} j_{\mu}-s(\{g_{\mu}\})
\end{equation}
where $e(\{g_{\mu}\})=\frac{\langle \mathcal H 
\rangle}{N/\rho} $ is the energy of a unit cell which is readily evaluated
using directly \eqref{Hamiltonian} and \eqref{corr}  
on a state specified by the set of correlations $\{g_{\mu}\}$.
The minimization of the above expression yields the inversion formulas:
\begin{equation}
 j_{\mu}=-\frac{\partial s(\{g_{\mu}\})}{\partial g_{\mu}}\label{inversion}
\end{equation}

We may state now our main result for the entropy $s(\{g_{\mu}\})$, which 
is given by 
\begin{equation}
 s(\{g_{\mu}\})=s^{(R)}(\{g_{\mu}\})-s^{(R-\rho)}(\{g_{\mu}\}). \label{entropy}
\end{equation}
The entropy \eqref{entropy} is written in terms of the functions 
$s^{(Q)}(\{g_{\mu}\})$ (to which we may
refer as ``entropy at range $Q$''), given by
\begin{align}
 s^{(Q)}(\{g_{\mu}\})=-\sum_{\tau_{Q}} p(\tau_{Q}) \log p( \tau_{Q}), 
\label{satrange1}\\
 p(\tau_{Q})=2^{-Q}\sum_{\mathrm{Rg}(\mu)\le Q} g_{\mu} O_{\mu} (
\tau_{Q}) \label{satrange2}
\end{align}
where $\tau_Q=\{t_1, t_2, \ldots, t_Q\}$ is the configuration of $Q$ auxiliary
Ising spins. Notice that the sum over $\mu$ now includes every subset,
including the null one.  The entropy can be shown to be convex in the variables
$\{g_{\mu}\}$, thus the equation \eqref{inversion} admits a solution, unless
some of the $p_Q$'s used in the calculation become negative, signaling a set of
``nonphysical'' correlations.

We now discuss the derivation of the formula \eqref{entropy}.
Let us think how the measurement of a correlation $g_\mu$ is operatively
defined: we look at $R$ consecutive spins
and we perform the measurement. Each of the microscopic configuration $\tau_R$
will occur with a given probability
$p(\tau_R)$ which would give rise to a mean value of $g_\mu$ given by
\begin{equation}
 g_\mu=2^{-R}\sum_{\tau_R} p(\tau_R) O_{\mu} (\tau_R).
\end{equation}
Since we know all of the correlations within the subsets of the $R$ spins, 
the system of the equations above may be inverted giving rise to 
\eqref{satrange2} with $Q=R$.
Then the Boltzmann formula $s=-\sum_i p_i \log p_i$ is applied to this set of
probabilities obtaining the expression $\eqref{satrange2}$ (always for $Q=R$). 
To derive \eqref{entropy} we calculate the entropy of the unit cell of
size $\rho$, regardless of the state of the remaining $R-\rho$  spins; in terms
of number of states it is 
\begin{equation}
 \sharp(\rho \;\text {spins})=\sharp(R \;\text {spins})/\sharp(R-\rho \; \text
{spins}) \label{numberofstates}
\end{equation}
where $\sharp(n \; \text{spins})$ denotes the number of microstates of a set of spins $n$
(subject to the constraints imposed by the correlations).
It should be noted that the $R-\rho$ spins to be traced out cannot be 
chosen at will: by inspection it turns out that the first $R-\rho$ spins 
is a good choice. Thus taking the logarithm of \eqref{numberofstates} 
we obtain our expression for the
entropy of a state characterized by the set of correlations $\{g_\mu\}$.
The number of correlations required to specify the state can be shown
by simple counting to be equal to $2^R-2^{R-\rho}$.

The above procedure can be formally applied also to a finite system of size 
$R$: this is achieved by letting $\rho=R$. 
In this case the system looks like a set of $N/R$
disjoint assemblies of $R$ spins for which the entropy is given by
$s^{(R)}(\{g_{\mu}\})$ being $s^{(0)}(\{g_{\mu}\})=0$. This result 
refers to a general finite system of Ising spins with arbitrary heterogeneous 
couplings without translational invariance
and it should be used if one wants to treat
datasets obtained from finite heterogeneous systems 
and extract Ising couplings \cite{mastromatteo2011}.
In general, the number of needed  
correlations to be known grows exponentially with the system size: 
in the case of the present paper, exponentially with $R$. Therefore, 
our result might be of practical importance if $R$ is small, $N$ is very large 
(\emph {i.e.}, near the thermodynamical limit) and the underlying 
Ising model is supposed to be one-dimensional.

Although the number of required correlation grows exponentially with
the range $R$, some simplifications may occur. For example if the system 
is known to be invariant under reflection $s_i\rightarrow-s_i$, then 
odd couplings vanish and we do not have
to measure the corresponding correlation function which are trivially zero. 
More generally, if we know that a coupling is equal to a given value
$j_{\nu}=j_{\nu}^0$ then one has the additional (nonlinear) equation:
\begin{equation}
 j_{\nu}^0=-\frac{\partial s(\{g_{\mu}\})}{\partial g_{\nu}},
\end{equation}
thus reducing the number of independent correlation functions.
The technique described may be easily adapted to other discrete spin systems
as Potts or Blume-Emery-Griffith models by using two or more Ising spins
to encode the state of the discrete spin, although such a mapping 
may obscure the symmetries of the original model. 

% The above problem bears some clear resemblance
% of with the one of a Markov chain with finite memory,
% such a mapping indeed is feasible and it is carried out
% in the Appendix \ref{markov}.
% 

To conclude this 
section, we finally observe that it is possibler 
to show in general the equivalence between the problem we have solved 
with the problem of finding, from a set of known correlations, 
the transition rates of an equilibrium 
- i.e. satisfying detailed balance - Markov chain with finite memory: we postpone such 
discussion to the Conclusions.

\section{Simple examples}\label{simple_examples}

\begin{figure}[t]
\begin{minipage}[b]{0.45\linewidth}
\centering
\includegraphics[angle=270, width=\textwidth]{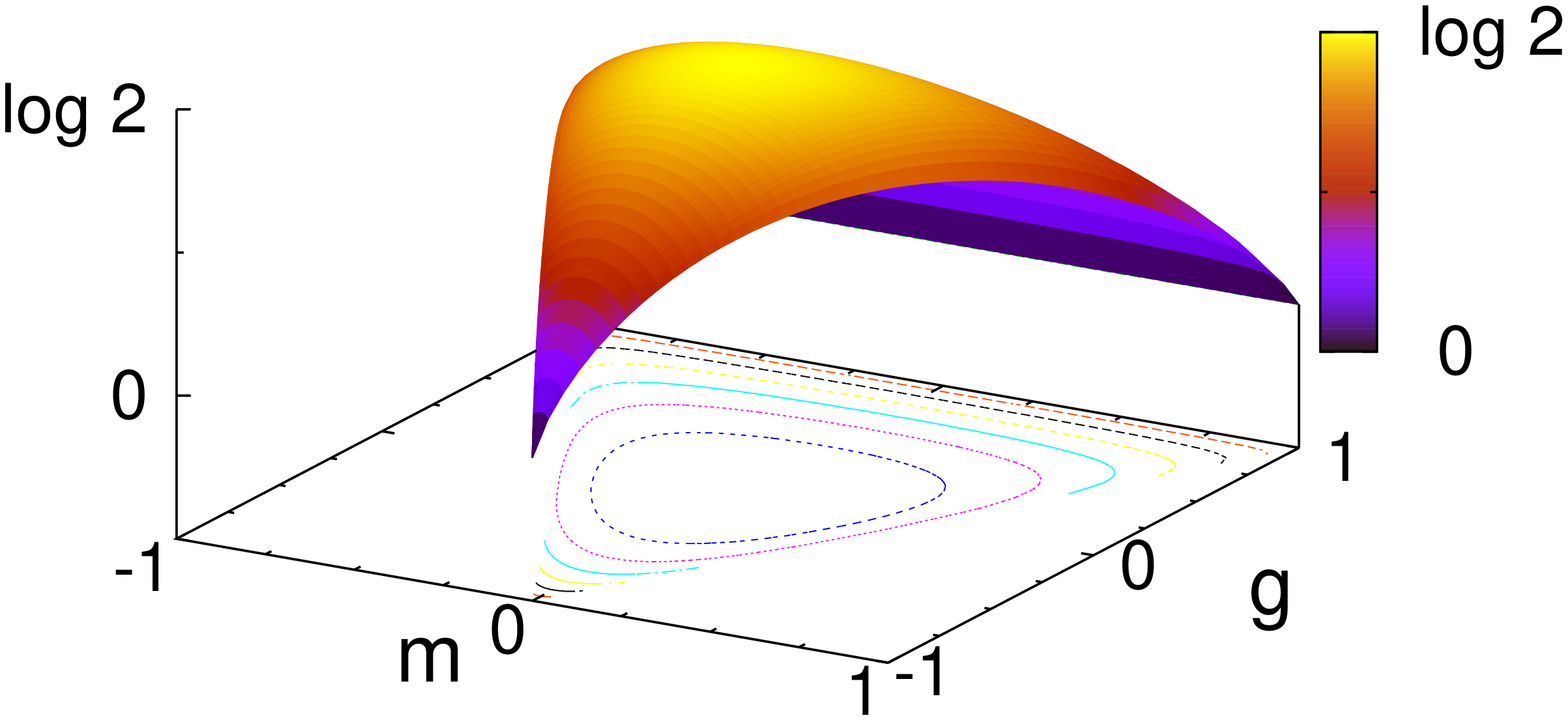}
\caption{Entropy per spin in terms of the nearest-neighbour correlation $g$
and the magnetization $m$ for $R=2$ and $\rho=1$.}
\label{entropy_g_m}
\end{minipage}
\hspace{0.5cm}
\begin{minipage}[b]{0.45\linewidth}
\centering
\includegraphics[width=\textwidth]{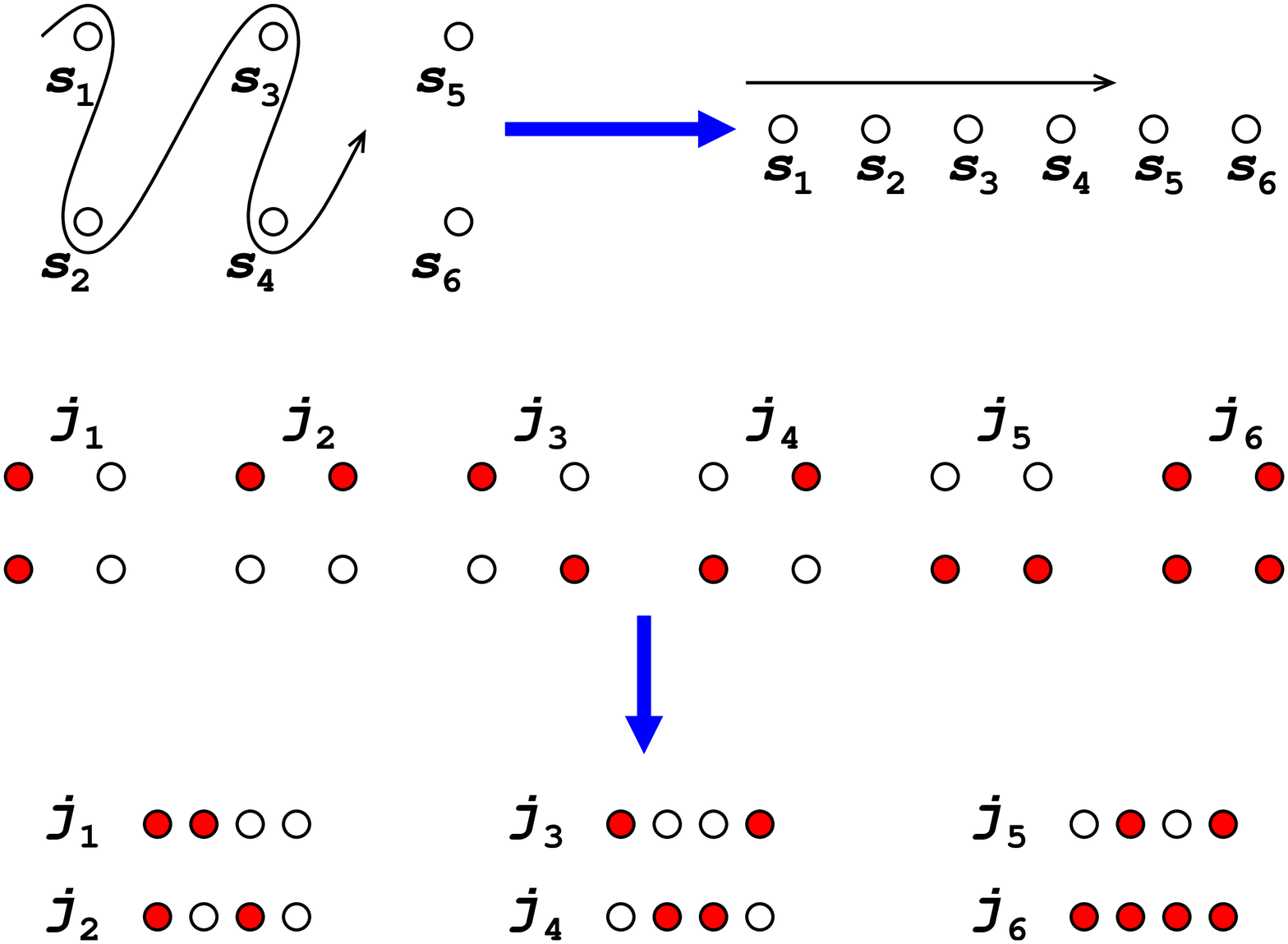}
\caption{Simple representation of how a spin ladder having 
$2$ legs may be mapped onto a spin chain with $R=4$ and $\rho=2$.
The filled circles represent spins which are present 
in the different operators considered (and associated to 
the couplings $j_1, \cdots, j_6$). In the middle part of the figure we represent them
on the ladder system, while on the bottom part we show how they look like 
on the chain. Explicitly these operators
correspond to the following terms in the Hamiltonian \eqref{HamiltonianExplicit}: $-j_1\sum_{i\;even} s_i s_{i+1}$, $-j_2\sum_{i\;even} s_i s_{i+2}$, 
$-j_3\sum_{i\;even} s_i s_{i+3}$, $-j_4\sum_{i\;odd} s_i s_{i+1}$, $-j_5\sum_{i\;odd} s_i s_{i+2}$ and $-j_6\sum_{i\;even} s_i s_{i+1} s_{i+2} s_{i+3}$}
\label{ladder_to_chain}
\end{minipage}
\end{figure}

As a first application of the results presented in the previous section,  
we consider a model with $R=2$ and $\rho=1$ (i.e. the Hamiltonian is
$\mathcal H=-h \sum_i s_i - j \sum_i s_i s_{i+1}$
where $h$ is the magnetic field and $j$ is the coupling). The only
independent correlations are the one-body correlator, 
i.e. the magnetization $m\equiv g_{\{1\}}$,
and the nearest-neighbour correlator $g\equiv g_{\{1,2\}}$.
Using \eqref{entropy} the entropy is calculated as
\begin{align}\label{s_m_g}
s(m,g)=-\frac{1+2m+g}{4} \log\left(\frac{1+2m+g}{4}\right)-\frac{1-2m+g}{4}
\log\left(\frac{1-2m+g}{4}\right)\nonumber\\-\frac{1-g}{2}
\log\left(\frac{1-g}{4}\right)+\frac{1+m}{2}
\log\left(\frac{1+m}{2}\right)+\frac{1-m}{2} \log\left(\frac{1-m}{2}\right)
\end{align}
which agrees with the expression obtained in \cite{mukamel2005} 
by combinatorial means.
In figure \ref{entropy_g_m} we plot the entropy: the 
convexity of $s$ guarantees to obtain the field and 
nearest-neighbour interaction in terms of
$m$ and $g$. The system we have just described presents no phase transitions,
apart from the zero temperature ones which occur at the border of the surface 
depicted in figure \ref{entropy_g_m}; in section \ref{MF_LR} we will see how 
the addition of a mean-field type interaction is easily included,
making phase transitions possible.
Interestingly on the lines $1\pm 2 m + g=0$ the system is frustrated
and our approach readily provides an expression for the 
ground state degeneracy. Differentiation of the entropy \eqref{s_m_g}
allows us to obtain the couplings, field $h\equiv j_{\{1\}}$ and
nearest-neighbour interaction $j\equiv j_{\{1,2\}}$ conjugated to
$m$ and $g$ respectively:
\begin{align}\label{h_and_j}
 h=&-\frac{\partial{s(m,g)}}{\partial m}=\frac{1}{2} \log \frac{(1-m)(1+2m+g)}{(1+m)(1-2m+g)}\\
 j=&-\frac{\partial{s(m,g)}}{\partial g}=\frac{1}{4} \log \frac{(1+2m+g)(1-2m+g)}{(1-g)^2}.
\end{align}
In Appendix \ref{transfer_matrices} we examine this example ($R=2$ and $\rho=1$)
and higher range ones ($R=3,4$ and $\rho=1$), explicitly checking the validity
of the inversion formula \eqref{inversion} using the transfer matrix method.

We will consider now a translationally invariant spin ladder with interaction
among the nearest-neighbors of the same and other chain.
For simplicity we will restrict ourselves to even interactions,
i.e. in the Hamiltonian only terms containing 
an even number of spins enter.
As shown explicitly in figure \ref{ladder_to_chain}
this system may be mapped onto a chain system with $R=4$ and $\rho=2$,
where the original interactions (allowed by symmetry)
and the new ones are shown. In terms of our
subset notation used in \eqref{Hamiltonian} the interaction parameters are defined as
$j_1\equiv j_{\{1,2\}}, j_2\equiv
j_{\{1,3\}}, j_3\equiv j_{\{1,4\}}, j_4\equiv j_{\{2,3\}}, j_5\equiv
j_{\{2,4\}}, j_6\equiv j_{\{1,2,3,4\}}$. This is easily generalized
to ladders made up of more than two chains and higher interaction range,
and thus our method is suited to treat general finite-range ladder systems.

We point out that the inversion formula allows to explicitly
write the relation 
among the $j$'s and $g$'s while the transfer matrix approach,
e.g. in the
simple ladder system described above,
already entails the solution of a fourth order
algebraic equation.

\subsection{Nearest-neighbour and next-to-nearest-neighbour plus four-spin interactions}
\label{comparison}
In this section we consider another simple example, in which nearest-neighbour and next-to-nearest-neighbour 
interactions are present together with a four-spin interaction: denoting 
the coefficients $j_{i,i+1}^{(2)}$, $j_{i,i+2}^{(2)}$ and $j_{i,i+1,i+2, i+3}^{(4)}$ by $j_1$, $j_2$ and $\lambda$ respectively,
the Hamiltonian \eqref{HamiltonianExplicit} reads
\begin{equation}
 \mathcal H =-\sum_{i} \left( j_{1} s_{i} s_{i+1}+j_2 s_{i} s_{i+2}+\lambda s_i s_{i+1} s_{i+2}  s_{i+3} \right).
\label{HamiltonianExplicit_4}
\end{equation}
This Hamiltonian can be exactly treated in our framework; the case $\lambda=\;0$ (the $j_1-j_2$ model) is discussed in the Appendix.
Here we aim at comparing approximate inverse Ising methods against exact results, focusing in particular on the low-correlation expansion (LCE) \cite{sessak2009} 
which is in the following compared with exact findings. We will use the LCE discussed in \cite{sessak2009} using as input a finite number of correlations, coherently
with what is done in this work (notice that for the present case our method needs just four correlation functions in order to recover
exactly the couplings); the maximal
range of two-body correlators for the LCE is denoted by $R_{rec}$. 

\begin{figure}
\begin{center}
\includegraphics[angle=270, width=.49\textwidth]{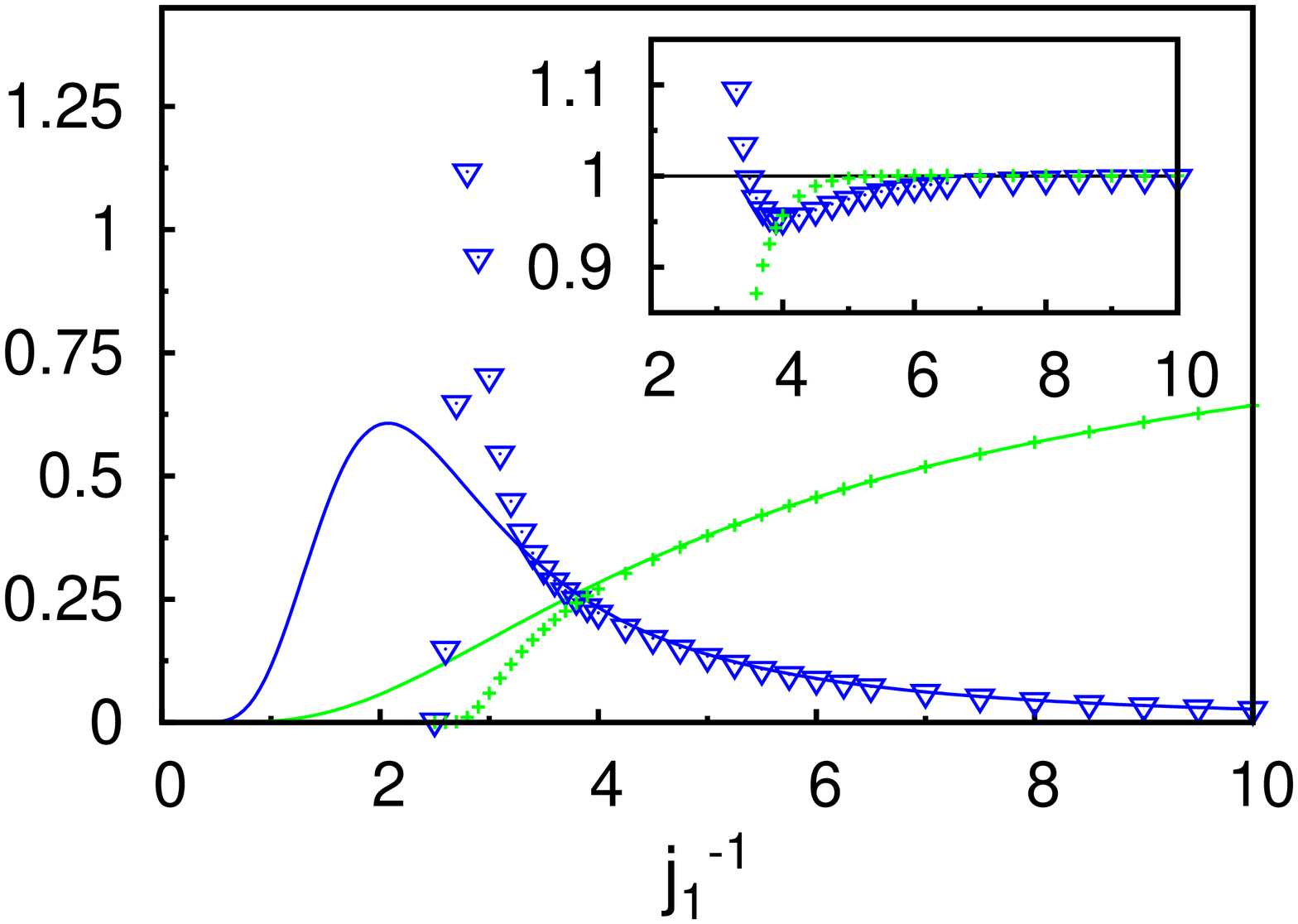}
\includegraphics[angle=270, width=.49\textwidth]{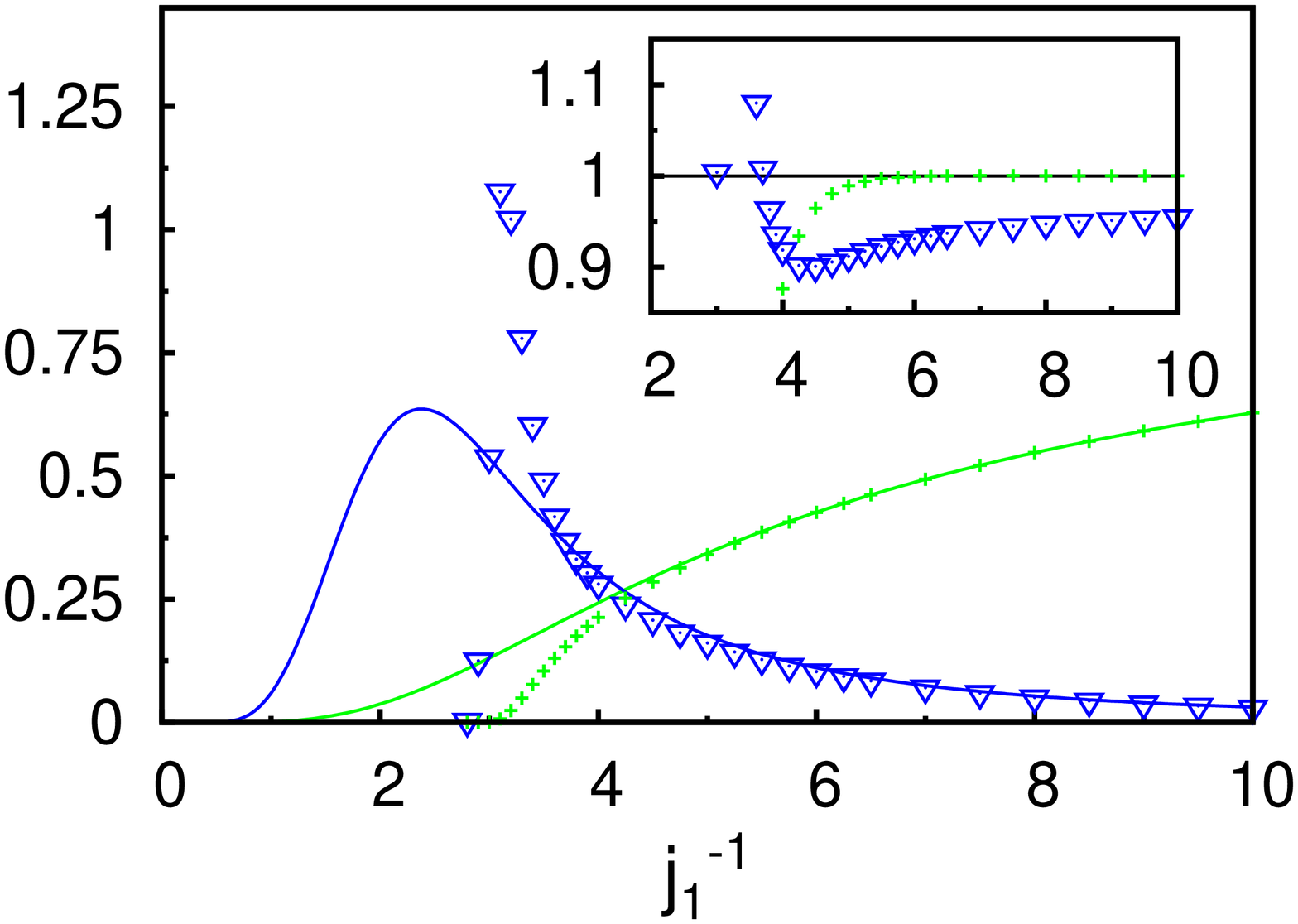}
\caption{Values for inverse susceptibility (green crosses) and specific heat (blue triangles) calculated 
with the LCE reconstructed couplings compared with the exact ones (full lines) as the inverse 
coupling $j_1$ is varied. The value $j_2/j_1$ is held fixed at the value $1$. The
figure on the right includes the four-spin interaction whose coupling is fixed at 
the value $\lambda=0.2\; j_1$. The inset shows the ratio between the predicted
values of specific heat and inverse susceptibility and the exact ones. 
Both figures are obtained keeping $R_{rec}=8$ correlation functions 
and using third order loop resummed expansion to reconstruct the couplings, as presented 
in \cite{sessak2009}.}\label{cV_chi}
\end{center}
\end{figure}

The LCE is discussed in \cite{sessak2009} to present an approximate technique for inverse Ising models having at most two-spin interactions: 
since Hamiltonian \eqref{HamiltonianExplicit} has only two-spin interactions for $\lambda=0$, we present LCE results for the case $\lambda=0$ in the Appendix 
where we discuss the $j_1-j_2$ model in detail, presenting the explicit solution using the transfer matrix approach. 
As expected, for low temperatures (i.e. large couplings),
the LCE breaks down and, as it can be seen in the right
panel of figure \ref{j1j2SM}, for moderate temperatures
the expansion may settle to an incorrect value of the
coupling as the range $R_{rec}$ is increased.
In order to further test the performance of the LCE against exact results we present 
in figure \ref{cV_chi} results for $\lambda=0$ and $\lambda=0.2 j_1$: although the LCE is developed in \cite{sessak2009} for two-spin interactions (and the extension to treat multispin interactions is expected to be cumbersome), the LCE reconstructed two-spin couplings with $\lambda \neq 0$ may partially take into account the effect of the four-spin interaction. To test to what extent this occurs, we consider 
two observables, susceptibility and specific heat, calculated
using the reconstructed couplings: the comparison 
with the exact results is in figure \ref{cV_chi}. As we can
see in the left panel the specific heat is more sensitive than
the susceptibility to reconstruction errors, even without four-spin interaction ({\emph i.e.} $\lambda=0$). This may be traced 
back to this type of LCE inversion procedure which aims
at reproducing two body correlators which in this model are 
required to calculate the susceptibility while the specific heat
already contains averages of four body
operators. 
Obviously the LCE, being not 
designed to infer models with multispin interaction, gives no hint
on the value $\lambda$ but in the right panel of figure \ref{cV_chi} we apply it for $\lambda=0.2$ in order to test how it can reproduce 
the considered  observables anyway: we can see that the
addition of such an operator reduce the temperature range
where the observables are correctly
reproduced. As noted above the specific
heat is more subjected to errors than the susceptibility
at contains higher-body operators.
From numerical inspection we saw that the LCE rather well performs in the high temperature regime even for relatively large values 
of $\lambda$, but deviates from exact results at lower temperature even for small values of $\lambda$ as shown in figure \ref{cV_chi}. 

\section{Exponential and power-law two-spin plus higher-spin couplings}\label{realistic_examples}

\begin{figure}\label{reconstruction}
\begin{center}
\includegraphics[angle=270, width=.49\textwidth]{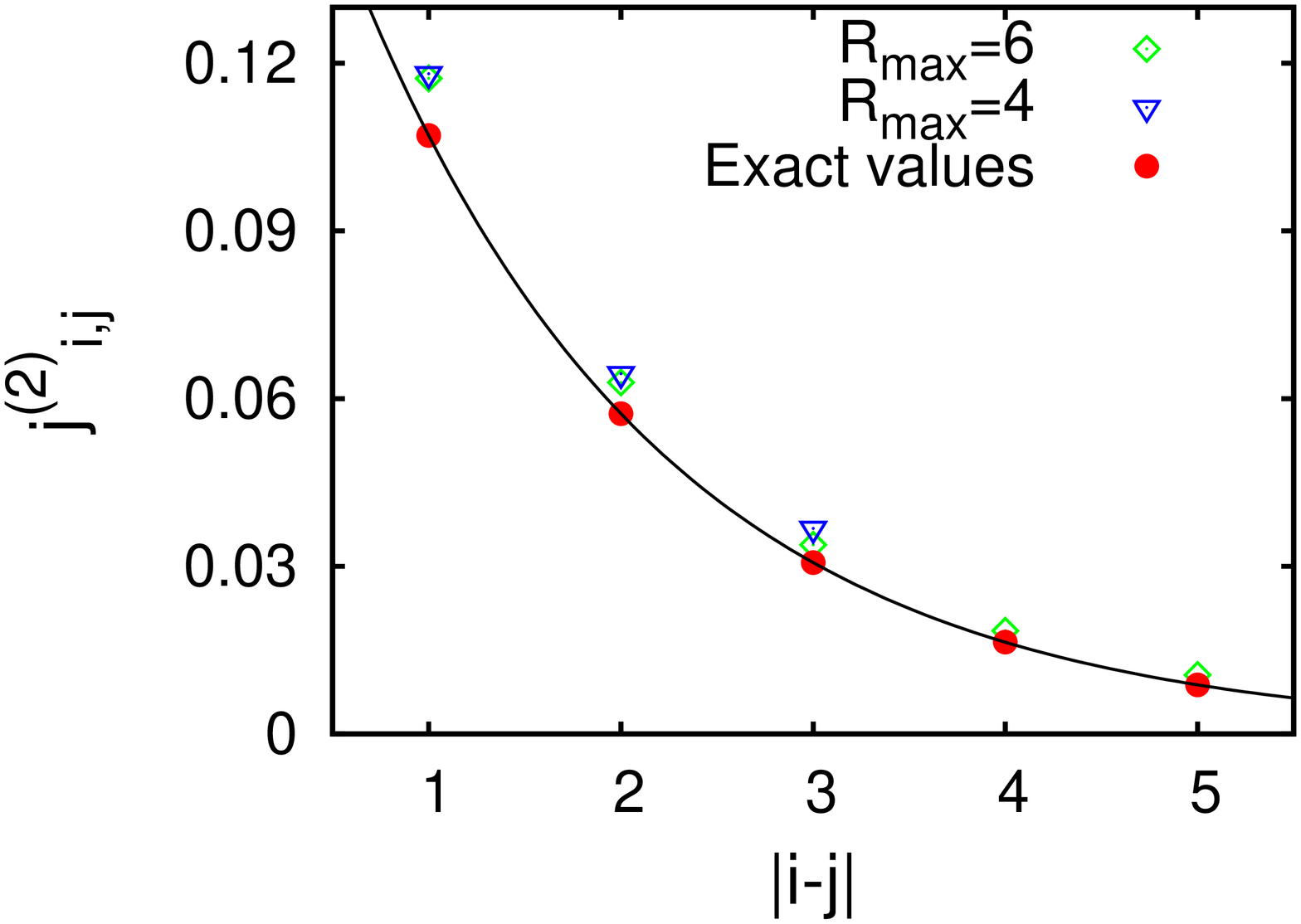}
\includegraphics[angle=270, width=.49\textwidth]{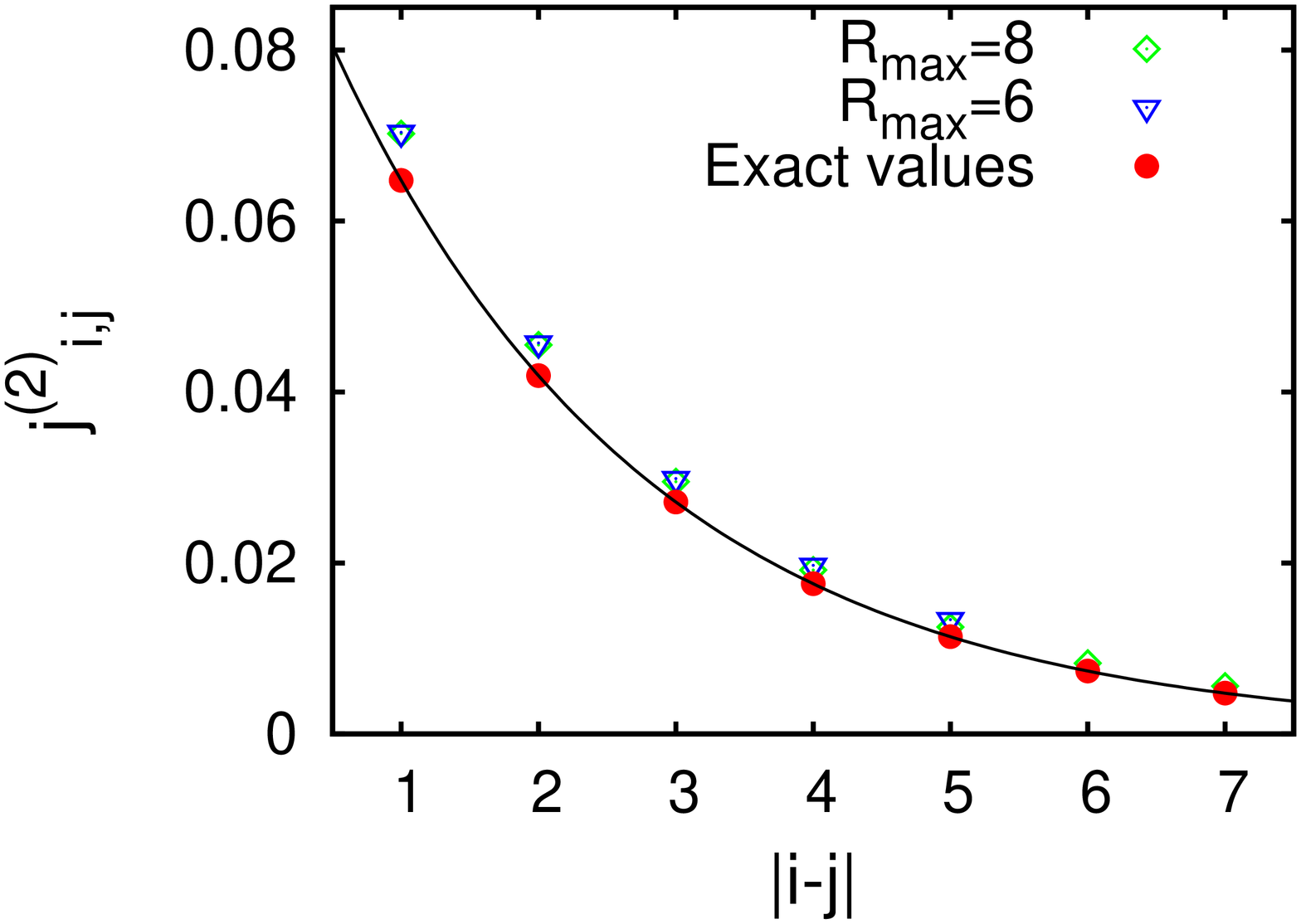}
\end{center}
\caption{Two examples of reconstructed values of the two-body couplings
$j^{(2)}_{i,j}$ (empty symbols) and the values of the couplings 
really used to generate the correlations (filled circles) - 
the full line is a guide to the eye. 
The figures refers to Hamiltonian \eqref{Hamiltonian_MCI} 
with parameters $J_0=0.2$, $\xi=1.6$, $R_{max}=6,4$ (left)
and $J_0=0.1$, $\xi=2.3$, $R_{max}=8,6$ (right).}
\end{figure}

\begin{figure}
\begin{center}
\includegraphics[angle=270,width=.8\textwidth]{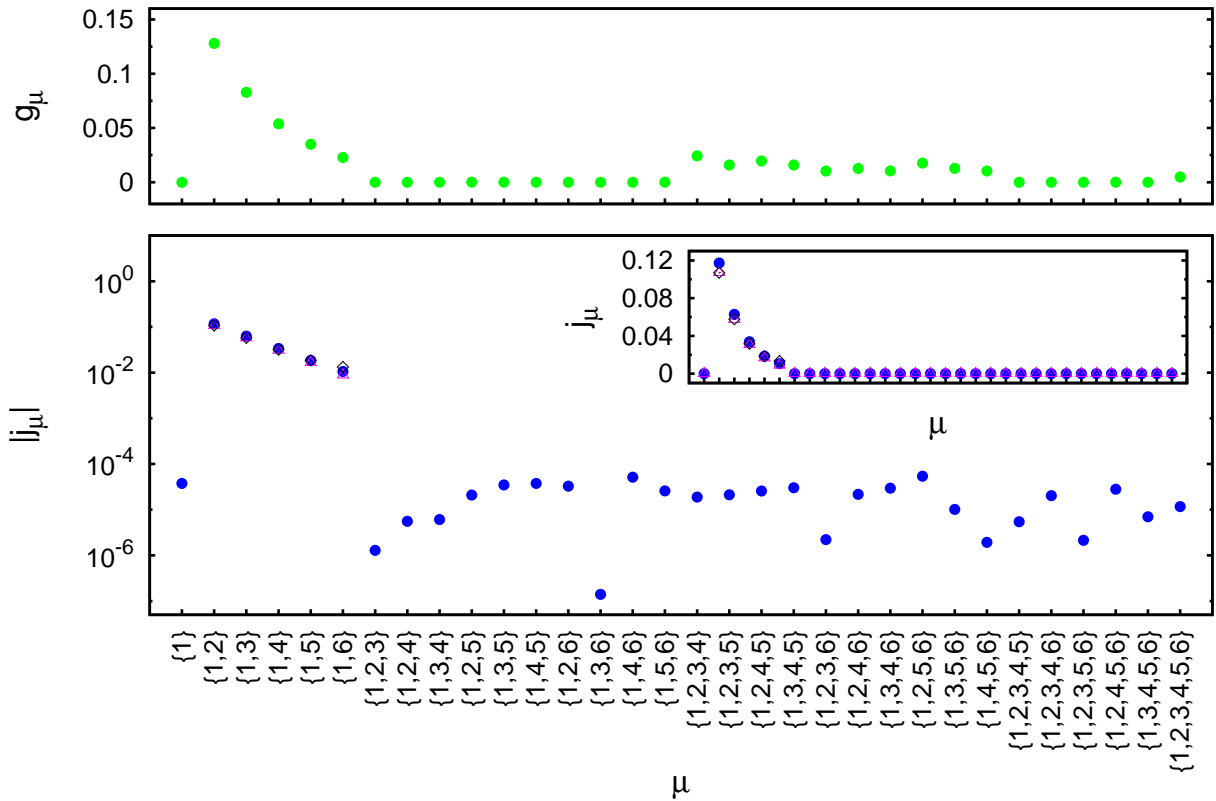}
\caption{Measured values of the correlations $g_\mu$ (top) and the inferred
couplings $j_\mu$ (bottom) for the $\mu$'s allowed by 
translational symmetry (filled blue circles) and exact values (empty purple triangles). 
The black diamonds refer to LCE results obtained with the perturbative expansion
up to third order \cite{sessak2009}. The figure at the bottom is a logarithmic
plot of the absolute values of
the $j_\mu$'s, while the inset gives for comparison the linear 
plot of the same 
$j_\mu$'s. The figures refer
to Hamiltonian \eqref{Hamiltonian_MCI} with parameters $J_0=0.2$, $\xi=1.6$. 
The reconstruction range is $R_{max}=6$. 
In abscissa are reported the $\mu$'s denoting the various coupling 
and correlators with the subset notation introduced 
in section \ref{notation_and_statement},
e.g. $g_{\{1\}}=\langle s_1\rangle$ is the correlation 
of the subset $\mu=\{1\}$,
$g_{\{1,2,4\}}=\langle s_1 s_2 s_4\rangle$ 
is the correlation 
of the subset $\mu=\{1, 2, 4\}$ and so on.}
\label{gj1}
\end{center}
\end{figure}

In this section we consider examples where we cannot access the 
full knowledge of our system: our inversion procedure will therefore yield
approximate results. First, we 
consider an Ising model with an exponentially decaying two-body 
interaction 
\begin{equation}\label{Hamiltonian_MCI}
H_I= -\sum_{(i,j)} j^{(2)}_{i,j} s_i s_j,\; \quad 
j^{(2)}_{i,j}=J_0 e^{-|i-j|/\xi}.
\end{equation}
Since the interaction now is not formally 
of ``finite range'' i.e. it does not vanish for distances beyond 
a given value of $R$, the transfer matrix method is not viable (although 
we still may perform a finite-range scaling in the size of the transfer matrix
\cite{uzelac1988,uzelac1989}). The set of synthetic
correlation functions is generated by a Monte Carlo method. 
Of course we  will not record all of the correlation function, 
but we will fix a maximal range $R_{max}$, 
thus we will have to measure on the order of $2^{R_{max}}$ 
correlation functions. The results for such a reconstruction are shown 
in figure \ref{reconstruction} for two values of the parameters. 
We see that the agreement improves as the
value of $R_{max}$ is increased (at the expense of calculating a larger number of correlation
functions). 
In figure \ref{gj1} a full set of the correlations and
inferred couplings is shown; if we look at the lower panel, 
the nonzero couplings are
clearly singled out (even for a value of $R_{max}$ as low as $6$), 
thus our reconstruction procedure
gives useful hints to build a faithful model of an unknown system. 
In figure \ref{gj1} results obtained with the LCE are also reported: one sees that there is a good agreement for the considered 
value of the temperature between the LCE results and the findings obtained using our reconstruction procedure.

\begin{figure}
\begin{center}
\includegraphics[angle=270,width=.8\textwidth]{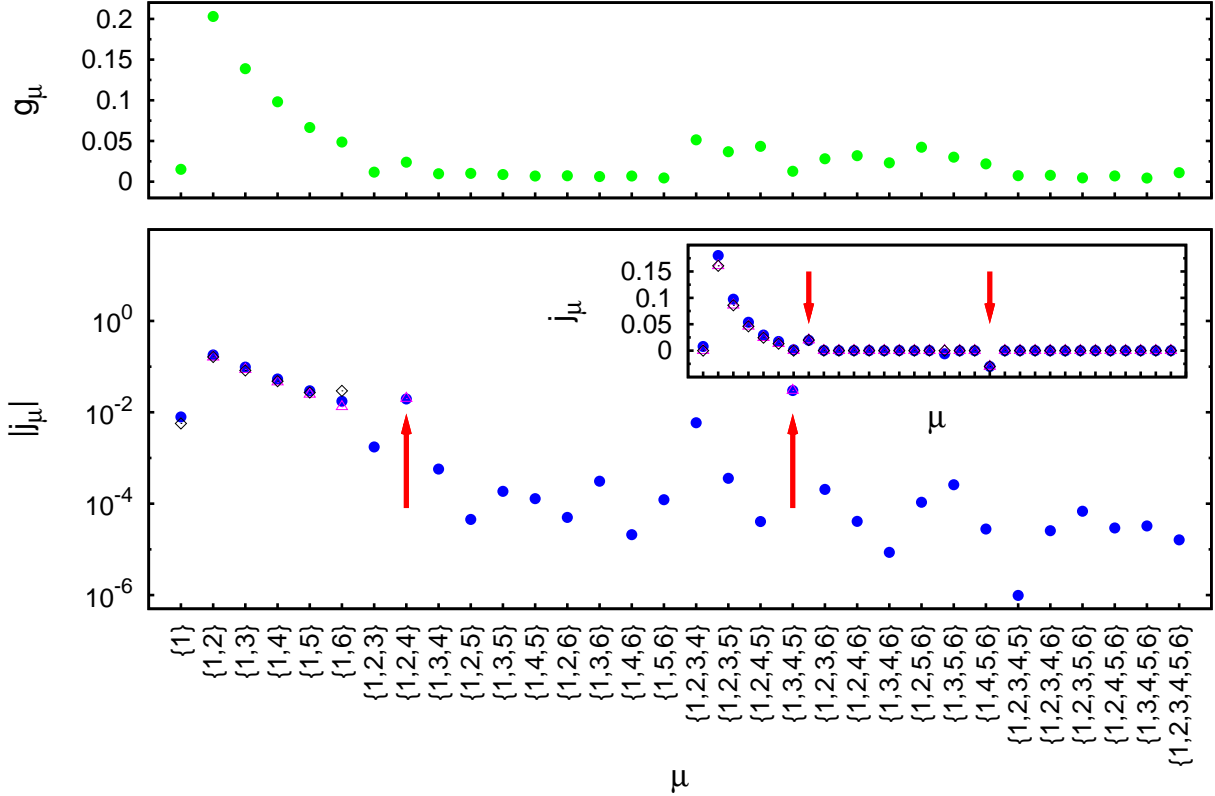}
\caption{Same as figure \ref{gj1} except for the black diamonds which are obtained 
with the TAP approach \cite{roudi2009}. The figures refer
to Hamiltonian \eqref{Hamiltonian_MCII} with parameters
$J_0=0.3$, $\xi=1.6$, $j_{\{1,2,4\}}=0.02$, $j_{\{1,3,4,5\}}=-0.03$. The reconstruction range 
is $R_{max}=6$. The arrows mark the multispin interactions.}
\label{gj2}
\end{center}
\end{figure}

\begin{figure}
\begin{center}
\includegraphics[angle=270,width=.8\textwidth]{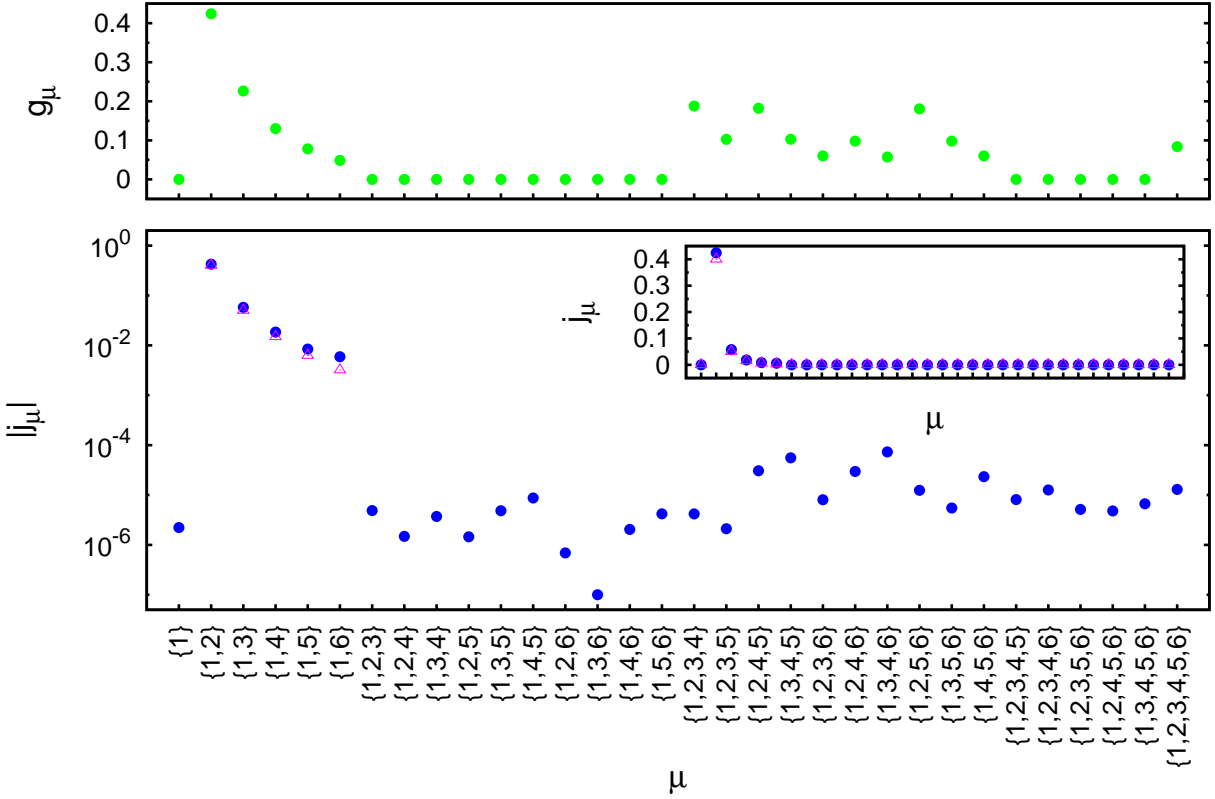}
\caption{Same as figure \ref{gj1}. Measured values of the correlations $g_\mu$ (top) and inferred
couplings $j_\mu$ (bottom) for the power-law decaying Hamiltonian \eqref{Hamiltonian_MCIII} 
with parameters $J_0=0.4$, $\alpha=3$. The reconstruction range is $R_{max}=6$.}
\label{gj3}
\end{center}
\end{figure}

In order to test the procedure on a system with more-body couplings 
we consider the Hamiltonian
\begin{equation}\label{Hamiltonian_MCII}
H_{II}= H_I - j_{\{1,2,4\}}\sum_i s_{i+1} s_{i+2} s_{i+4} - j_{\{1,3,4,5\}}\sum_i
s_{i+1} s_{i+3} s_{i+4} s_{i+5}
\end{equation}
which includes three- and four-body interactions. 
As we see in figure \ref{gj2}, 
even in this case the reconstruction procedure gives useful hints 
on the couplings present in the system, although some of the inferred couplings,
which were zero in the starting model,
are predicted to be of comparable size to the nonzero ones 
(especially the $j_{\{1,2,3,4\}}$ coupling).
This is due to the finite reconstruction range
and to the, albeit small (indeed smaller than the symbols 
in the figures \ref{gj1}, \ref{gj2}, \ref{gj3}),
errors in the determination of the correlation.
This implies that in order to clearly distinguish
the contribution of the different couplings, the correlation
should be known with high accuracy.
As for modeling purposes, this is not a problem since
the values of the coupling obtained give rise to a set 
of correlations not distinguishable
from the original one. For reference we also plot in figure \ref{gj2} the results of the TAP
equation approach developed in \cite{roudi2009}, which of course provides no
information on the multispin couplings, but as far as one-body
operators and two-body operators are concerned this approach at the 
temperature considered in figure \ref{gj2}
performs very well.

Finally we examine a model with power-law decay of the interaction
\begin{equation}\label{Hamiltonian_MCIII}
H_{III}=-\sum_{(i,j)} j^{(2)}_{i,j} s_i s_j,\; \quad j^{(2)}_{i,j}=
\frac{J_0}{|i-j|^\alpha}.
\end{equation}
As can be seen in figure \ref{gj3} (with $\alpha=3$), 
the results are good also in this case: it 
can be generally observed that the 
reconstructed interactions are higher than the exact ones,
due to the fact that the interaction within the reconstruction range
have to account for the interactions lying outside this range.

\section{Mean-field couplings}\label{MF_LR}

In this section we briefly discuss 
how our previous results can be used in presence of
mean-field long-range interactions, showing that our inversion approach 
may be used on this class of systems. 
We consider a system with energy $e$ of general form, i.e.
a non-linear function of the correlators. The number 
of couplings entering
the energy $e$ should still equal the number of independent correlation
functions in order to perform, at least in principle, the inversion procedure. 
Such an energy will be denoted by $e^{MF}(\{g_{\mu}\},\{j^{MF}_m\})$
where the index $m$ runs over the mean-field couplings.

By requiring the free energy 
$f^{MF}(\{g_{\mu}\},\{j^{MF}_m\})=e^{MF}(\{g_{\mu}\},\{j^{MF}_m\}) - s(\{g_{\mu}\})$
to have a minimum when the ${g_\mu}$ are set to the known
values will give the equations implicitly determining 
all the couplings, including the mean-filed ones 
$\{j^{MF}_m\}$. When the energy is differentiable
these equations read
\begin{equation}\label{inversionMF}
 \frac{\partial e^{MF}(\{g_{\mu}\},\{j^{MF}_m\})}{\partial g_{\mu}}=
\frac{\partial s(\{g_{\mu}\})}{\partial g_{\mu}}.
\end{equation}
This set of equations will in general have multiple 
solutions or possibly no solutions at all. If more solutions
are present the one (or ones) rendering the free energy minimal
should be chosen. The points where the absolute minimum
of the free energy branches or it changes discontinuously will signal 
a phase transition. As described in the section \ref{inversion_formula}
some values of the couplings may be known in advance thus
reducing the number of equations to be solved. Another 
possibility is that a function of the coupling is fixed;
a notable example being the energy itself, corresponding
to the microcanonical description of the system.
We remark that in the class of models we consider
all of these steps can be carried out exactly
since the explicit form of entropy \eqref{entropy} is known.

As an example we may consider the first model examined
in section \ref{simple_examples} ($R=2$, $\rho=1$) by adding 
mean-field two-body couplings.
Instead of the energy $e(m,g)=-h m-j g$ (where 
$h=j_{\{1\}}$ and $j=j_{\{1,2\}}$ ) we will
set $e=e^{MF}(m,g)=-j^{MF} m^2-j g$. 
The appearance of the nonlinear term $m^2$
in the energy is due to the presence of non-local mean-field 
operators in the spin Hamiltonian like $-j^{MF}/N 
\left(\sum_{i=1}^{N} s_i \right)^2$.
It should be noted that such an operator is not 
uniquely defined, e.g. the operator 
\begin{equation}
-\frac{j^{MF}}{\sum_{i=1}^{N} \frac{1}{i^{\alpha}}} 
\sum_{i,j} \frac{s_i s_j}{|i-j|^\alpha} \quad \left(0<\alpha<1\right) 
\end{equation}
and other Kac-rescaled nonextensive potentials give rise to the 
term $-j^{MF} m^2$ in the energy density at the thermodynamic
limit when evaluated on a state
with magnetization $m$ and nearest-neighbour 
correlation $g$ \cite{vollmayr-lee2001}.
If we set $j=0$ we get the usual mean-field model,
otherwise we obtain a model with competing mean-field 
and short-range coupling introduced by Kardar 
\cite{kardar1983}, exhibiting 
 a complex phase diagram which shows nonequivalence between 
the canonical and microcanonical description \cite{mukamel2005} 
(for a review on inequivalence between ensembles and other issues concerning
nonextensive systems see \cite{ruffo2009}).
As already observed the entropy of such a model within our approach 
is easily computed [see \eqref{s_m_g}] and it could be generalized
thus allowing one to treat one-dimensional models possessing 
multiple competing finite-range
and mean-field interactions.

\section{Conclusions}\label{conclusions_and_perspectives}

In this paper we presented the explicit 
solutions of the inverse Ising problem for a one-dimensional translational
invariant model with arbitrary finite-range 
multispin interactions once a number $\sim 2^R$ (where $R$ is the range of the 
interactions) of independent correlations is known. 
When applied to unknown systems this method
correctly detects arbitrary interactions;
our results are then applied to systems with a range extending 
beyond the one set by maximum distance of the spins of the 
recorded correlation functions, giving useful hints on the interactions that should be kept in an 
effective model. 

As an application, we reconstructed the couplings of chain Ising Hamiltonians having 
exponential or power-law two-spin plus three- or four-spin couplings.
We also discussed the generalization 
of the method to ladders. Mean-field interactions can be also included in the framework, 
allowing us to describe systems exhibiting phase transitions.
The presence of both finite-range (local) and mean-field (nonlocal) 
interactions can give
rise to interesting competition effects greatly enhancing the
descriptive power of the models we can exactly 
solve with our techniques. Our results provide then a theoretical laboratory where different 
approximate inverse Ising techniques can be benchmarked against the exact
results obtained using our method: in the paper we performed such comparison is some illustrative examples. 

The one-dimensional inverse Ising problem we have solved in the present paper is analogous to
what may be called the inverse Markov chain problem: given a specific set of correlations at equilibrium, 
find the corresponding transition rates.
The two inversion (Ising and Markov) problems are related since it is possible to associate an Ising model to an equilibrium 
Markov Ising chain with
finite memory in full generality: to show it, 
for definiteness let us consider Ising variables
(although extensions to other discrete state spaces
is straightforward). The finite-range $R$ in our solution 
of the inverse one-dimensional Ising problem is the counterpart of the finite memory in the inverse Markov problem: 
let the state of the next $\rho$ spins be ruled by the 
state of the preceding $R-\rho$ spins.
%Thus we are dealing with a Markov chain with finite memory.
We put the new spins on the right side of the old ones: the time of the Markov process is increasing from left to right. 
The correlations impose
constraints on the transition rates: it turns out that
the number of independent correlations required to solve
the inverse Markov problem is the same as the number needed to solve the (related) 
inverse Ising problem in one dimension of range $R$
and period $\rho$, i.e. the number of independent correlations is $2^R-2^{R-\rho}$.
Adapting the procedure discussed in section \ref{inversion_formula} which lead to \eqref{satrange2} it is possible to compute 
the transition rates from the state $\tau_{R-\rho}$ of $R-\rho$ spins 
to the state $\theta_{\rho}$ of $\rho$ spins. These transitions are given by
\begin{equation}
 w_{\tau_{R-\rho}\rightarrow\theta_{\rho}}=\frac{p(\eta_R)}{p(\tau_{R-\rho})},\label{trans_rates}
\end{equation}
where the $p$'s are calculated according to formula \eqref{satrange2}
where the input correlations $\{g_\mu\}$'s are plugged in, and
$\eta_R$ is the configuration of $R$ spins obtained by juxtaposing the
states $\tau_{R-\rho}$ (on the left) and $\theta_{\rho}$ (on the right).
The transition matrix obtained this way is a $2^{R-\rho}$ by $2^{\rho}$
matrix; in order to have a square matrix we have to fold more steps
of the the transition matrix until we obtain the probability to
go from a set of $\max (R-\rho,\rho)$ to the next $\max (R-\rho,\rho)$
spins \cite{gori_future}. The transition matrix satisfies detailed 
balance by construction, therefore this is a reversible Markov
chain. This mapping has already been worked out in a discrete different form
in \cite{vanderstraeten2009} where the connection to discrete statistical models
is also discussed. 

According to the previous discussion, we may thus associate an Ising model to the an equilibrium Markov Ising chain with
finite memory in full generality, allowing us to treat systems 
where one direction (typically time) is singled out. 
As a possible example deserving future investigation one could consider time-series of financial data and try to estimate 
with the procedure discussed before the transition probabilities of the associated guessed Markov chain in order 
to test the validity of such a description. It is intended - in order to apply the previous results - 
that the analyzed data should be discretized on a timescale 
such that nontrivial correlations occur, and that the whole time of observation 
is such that the system can be reliably considered at equilibrium. Obviously a way to encode significant information in an Ising
variable has to be devised, being this in general a nontrivial
task; for example, we may think of ``up'' and ``down'' spin corresponding to a price raise or
lowering respectively. The next step would be to analyze the correlations among different time-series of data to determine  
if and how correlations among different stocks occur. For example the Ising ladder system depicted 
in figure \ref{ladder_to_chain} may reproduce the correlations among two stocks whose state depends on the value of the other stock 
at the same timestep and on the value of the same or other stock at the previous timestep (for simplicity, in figure \ref{ladder_to_chain} 
odd interactions are not depicted). {\emph E.g.}, it should be noted that the inclusion of the interaction dubbed $j_6$ 
in figure \ref{ladder_to_chain} may reproduce some kind of nontrivial many-body interaction among the stocks. Extending the number of chains in the ladder system
and/or the range allows us to treat larger sets of stocks with longer correlations in time. 

We think that studying stationary time-series of correlated data using 
the techniques presented in this paper (and as well the mapping onto Markov chains discussed in this section) will be
an interesting subject of future research. 
In perspective, one could apply the method here discussed to 
datasets and/or statistical mechanics models 
which are supposed to be described by effective one-dimensional 
Ising chains near the thermodynamical limit. 
To this aim, one should address in the future a treatment
of the case where large errors in the measured correlation functions 
are present and/or some of the correlations are missing; our exact 
result could be a good starting point to move in that direction. Next, 
our results could be extended in higher dimensional cases (where of course
one does not expect to find closed formulas), hierarchical or 
tree-like models. Some preliminary results in the two dimensional case 
seem to indicate that this approach leads to equations resembling 
the Dobrushin-Lanford-Ruelle ones \cite{dobrushin1969,lanford1969}.
Another interesting direction could be to use the a renormalization group
approach on the correlation functions, in order to study how the 
couplings determined by correlations at some scale $R$ are related 
to the ones computed at a larger scale $R'$.

{\em Acknowlegments:} We wish to thank I. Mastromatteo and M. Marsili for many 
very useful discussions. This work has been supported by the grants INSTANS (from ESF)
and 2007JHLPEZ (from MIUR).

\appendix
\section{Transfer Matrices}\label{transfer_matrices}
In this Appendix we work out the transfer matrix 
for the general translational invariant ($\rho=1$) 
Ising model with range $R=2,3,4$ and we check that the 
results obtained analytically and numerically 
with the transfer matrix formalism are 
fully consistent with the predictions of the formula \eqref{entropy}.

We start by briefly recalling the method (see e.g. \cite{yeomans1992}). 
In general the transfer matrix $\mathbf{T}$ 
is built by identifying the $2^B$ states of a block of spins $B$ with
independent and orthogonal vector of a space of dimension $2^B$ 
such that the matrix elements of $\mathbf{T}$ are
\begin{equation}
\langle a |\mathbf{T}| b \rangle=e^{-\mathcal{H}_{int}(a)-\mathcal{H}_{ext}(a,b)},
\end{equation}
where $\mathcal{H}_{ext}(a,b)$ is the interaction energy among 
two consecutive blocks of spins $a$ and $b$ ($a$ is placed to the left of $b$), 
$\mathcal{H}_{int}(a)$ is the interaction energy among the spins 
belonging to the 
same block. The vector corresponding 
to the spin state $a$ is denoted, using the ket notation, by $|a\rangle$. The 
size of the blocks $B$ has to be chosen according to the range and size of the 
unit cell $\rho$ of the system in order to have all of the interaction terms 
contained in $\mathcal{H}_{int}$ or $\mathcal{H}_{ext}$. The partition function 
of the system of size $N$ is simply given by 
$\mathcal{Z}_N=\mathrm{Tr}(\mathbf{T}^{N/B})$ 
so that in the infinite size limit the free energy per unit cell may be written 
in terms of the largest eigenvalue $\lambda_{max}$ of $\mathbf{T}$:
\begin{equation}
 f=-\frac{1}{B/\rho}\log \lambda_{max}.
\end{equation}
The existence and unicity of $\lambda_{max}$ is guaranteed by Perron-Frobenius 
theorem, being all elements of $\mathbf{T}$ strictly positive (for nonvanishing
temperature). The correlation functions may be obtained just by 
differentiation of the free energy $f$.

We start with the simple example of $R=2$ 
which has been worked out in section \ref{simple_examples}.
The two independent couplings will be denoted as usual by 
$j_{\{1\}}=h$ (magnetic field) and 
$j_{\{1,2\}}=j$ (nearest-neighbour coupling). 
The $2$ by $2$ transfer matrix $\mathbf{T}$, with the
states identification $|\uparrow \rangle = (1,0)$ and 
$|\downarrow \rangle = (0,1)$, reads
\begin{equation}
\mathbf{T}=\left(\begin{array}{cc}
e^{h+j} & e^{h-j} \\
e^{-h-j} & e^{-h+j}
\end{array}\right).
\end{equation}
The free energy is then given by
\begin{equation}
 f(h,j)=-\log \left[ e^j \cosh h+ \sqrt{e^{2 j} \cosh^2 h- 2 \sinh (2 j)}\right].
\end{equation}
Differentiating $f(h,j)$ with respect to $h$ and $j$ 
gives the magnetization $m$ and the nearest-neighbour 
correlation $g$ respectively:
\begin{align}
 m=&-\frac{\partial{f(h,j)}}{\partial h}=\frac{e^j \sinh h}{\sqrt{e^{2 j} \cosh^2 h- 2 \sinh (2 j)}}\\
 g=&-\frac{\partial{f(h,j)}}{\partial j}=\coth (2 j)-\frac{\cosh h}{\sinh (2 j)\sqrt{1+e^{4j}\sinh^2 h}}.
\end{align}
The inversion of the above formulas 
with respect to $h$ and $j$ yields the expressions \eqref{h_and_j}.

\begin{figure}
\begin{center}
 \includegraphics[angle=270, width=.49\textwidth]{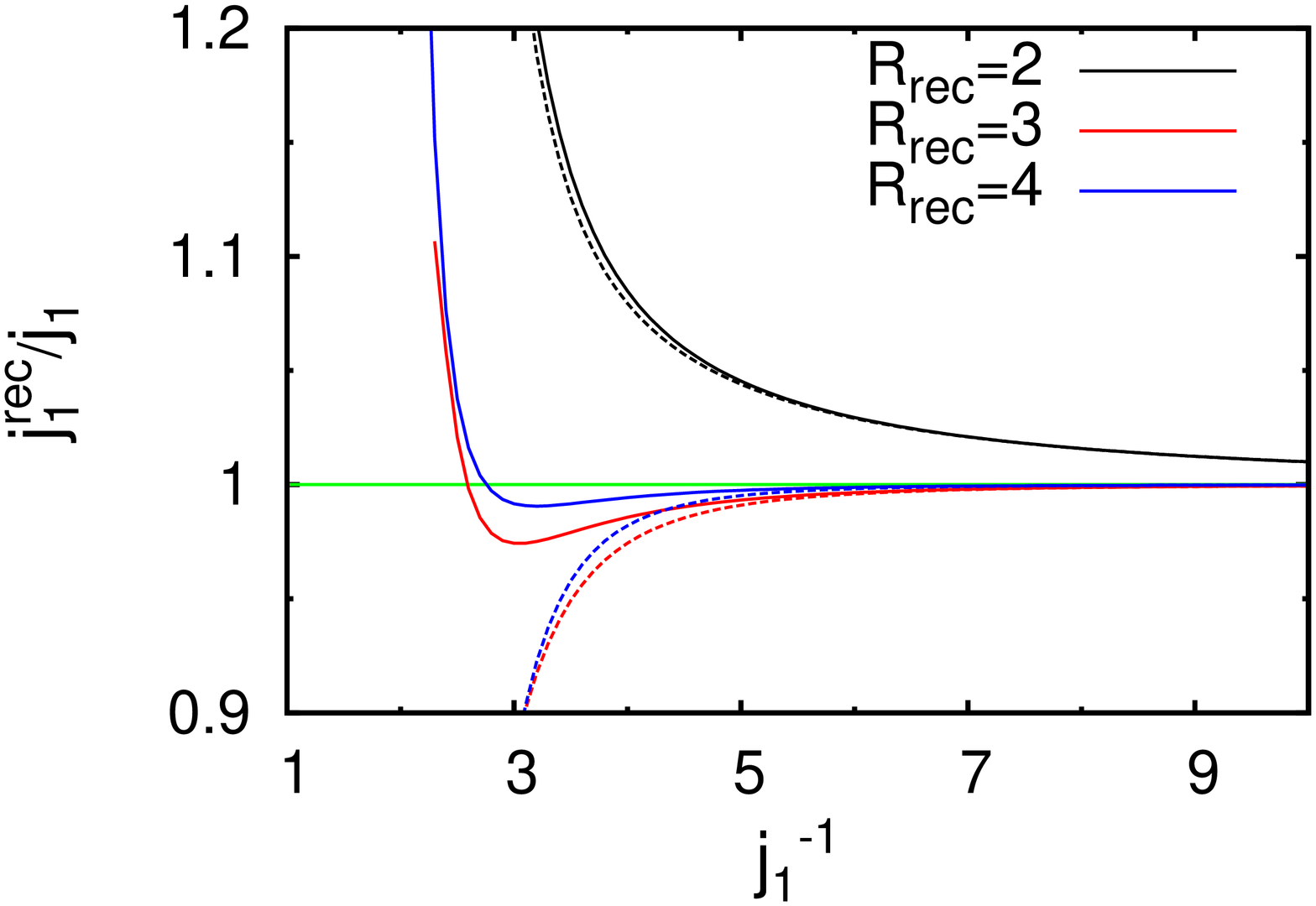}
 \includegraphics[angle=270, width=.49\textwidth]{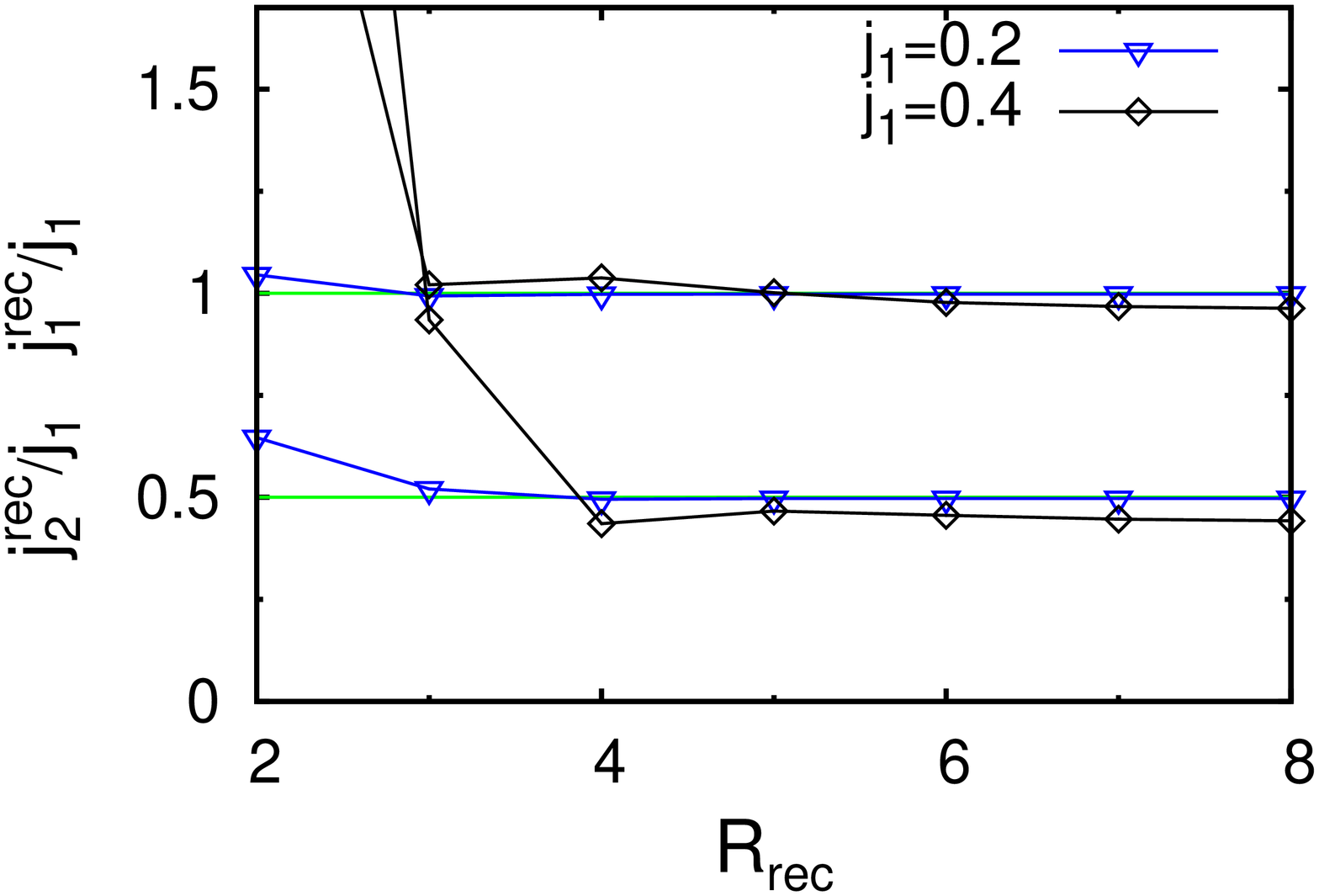}
\caption{On the left the ratio between the exact and reconstructed value of the nearest-neighbour
coupling in terms of $j_1$. The various lines denote how many correlation functions are kept in the reconstruction
formulas as indicated. The continuous and dotted line refer respectively to the third and seventh order
in the LCE of \cite{sessak2009} (including the loop contributions).
On the right the ratio of reconstructed $j^{rec}_1$ (above), $j^{rec}_2$ (below)
and the exact value of $j_1$ in terms of the reconstruction range $R_{rec}$ are 
shown. Only the result for the seventh order LCE is shown. In both figures the green 
straight lines are the exact values and the 
nearest-neighbour coupling $j_2$ is set to the half of $j_1$.}\label{j1j2SM}
\end{center}
\end{figure}

Now we consider a model with only even interactions and range $R=3$,
i.e. containing the coupling $j_1=j_{\{1,2\}}$ and $j_2=j_{\{1,3\}}$
which we dub $j_1-j_2$ model: the corresponding Hamiltonian is given by \eqref{HamiltonianExplicit_4} with $\lambda=0$. 
This model may be mapped onto the previous example ($R=2$, $\rho=1$)
by introducing ``kink'' variables $s_i s_{i+1}$
with the identifications $j_1=h$, $j_2=j$ (we do not report the corresponding results). 
In figure \ref{j1j2SM} we plot the couplings reconstructed according to the 
low-coupling expansion (LCE) introduced in \cite{sessak2009} against the exact results which can be found by the method discussed in this paper or by the transfer matrix approach.  
The LCE allows to infer the magnetic fields and the two-body couplings 
from the two-body correlators and magnetizations. 
In \cite{sessak2009} the LCE has been carried out, in the zero field case, up to seventh order in the correlations with
loop resummation. We have used the LCE as both the order
of the expansion and the number of correlation
function we assume to know are increased. The maximal range of two-body 
correlations which are used as input is denoted by $R_{rec}$. In figure \ref{j1j2SM} we depict the reconstructed
couplings $j_{1}^{rec}$, $j_{2}^{rec}$ and the exact ones for different
values of $R_{rec}$. The ratio $j_1/j_2$ is kept fixed,
thus $j_1$ serves as inverse temperature. As expected as the temperature
is lowered the agreement gets worse, and it may be noticed that
the inclusion of higher order terms in this case does not significantly 
improve the performance
of the inversion: as one can see, the lower order results depicted
in the left panel of figure \ref{j1j2SM} is more
reliable at lower temperatures (this is the reason why in figure \ref{cV_chi}
we employed the third order LCE).
In the right panel of figure \ref{j1j2SM} we can see how the 
the increase of $R_{rec}$ improves the 
quality of the inversion but beyond
a given range the reconstructed couplings $j_{1,2}^{rec}$
settle to a value 
which, in the lower temperature case
examined, deviates from the exact one.

We move on to the next example, the $R=3$ case with no restriction on the
symmetry of the couplings.
Identifying the states as 
$|\uparrow \uparrow \rangle = (1,0,0,0)$, 
$|\uparrow \downarrow \rangle = (0,1,0,0)$, 
$| \downarrow \uparrow \rangle = (0,0,1,0)$, 
$| \downarrow \downarrow \rangle = (0,0,0,1)$
and the couplings as 
$j_{\{1\}}=j_1$, $j_{\{1,2\}}=j_2$, $j_{\{1,3\}}=j_3$, $j_{\{1,2,3\}}=j_4$,
the transfer matrix reads
\begin{equation}
\mathbf{T}=\left(\begin{array}{cccc}
e^{2 j_1 + 2 j_2 + 2 j_3 + 2 j_4}& e^{2 j_1 - 2 j_4}& e^{2 j_1 + 2 j_2}& e^{2 j_1 - 2 j_3}\\
1& e^{-2 j_2 + 2 j_3}& e^{-2 j_3 - 2 j_4}& e^{-2 j_2 + 2 j_4}\\
e^{-2 j_2 - 2 j_4}& e^{-2 j_3 + 2 j_4}& e^{-2 j_2 + 2 j_3}& 1\\
e^{-2 j_1 - 2 j_3}& e^{-2 j_1 + 2 j_2}& e^{-2 j_1 + 2 j_4}& e^{-2 j_1 + 2 j_2 + 2 j_3 - 2 j_4}.
\end{array}
\right)\label{TM2}
\end{equation}

\begin{figure}
\begin{center}
\includegraphics[angle=270, width=.49\textwidth]{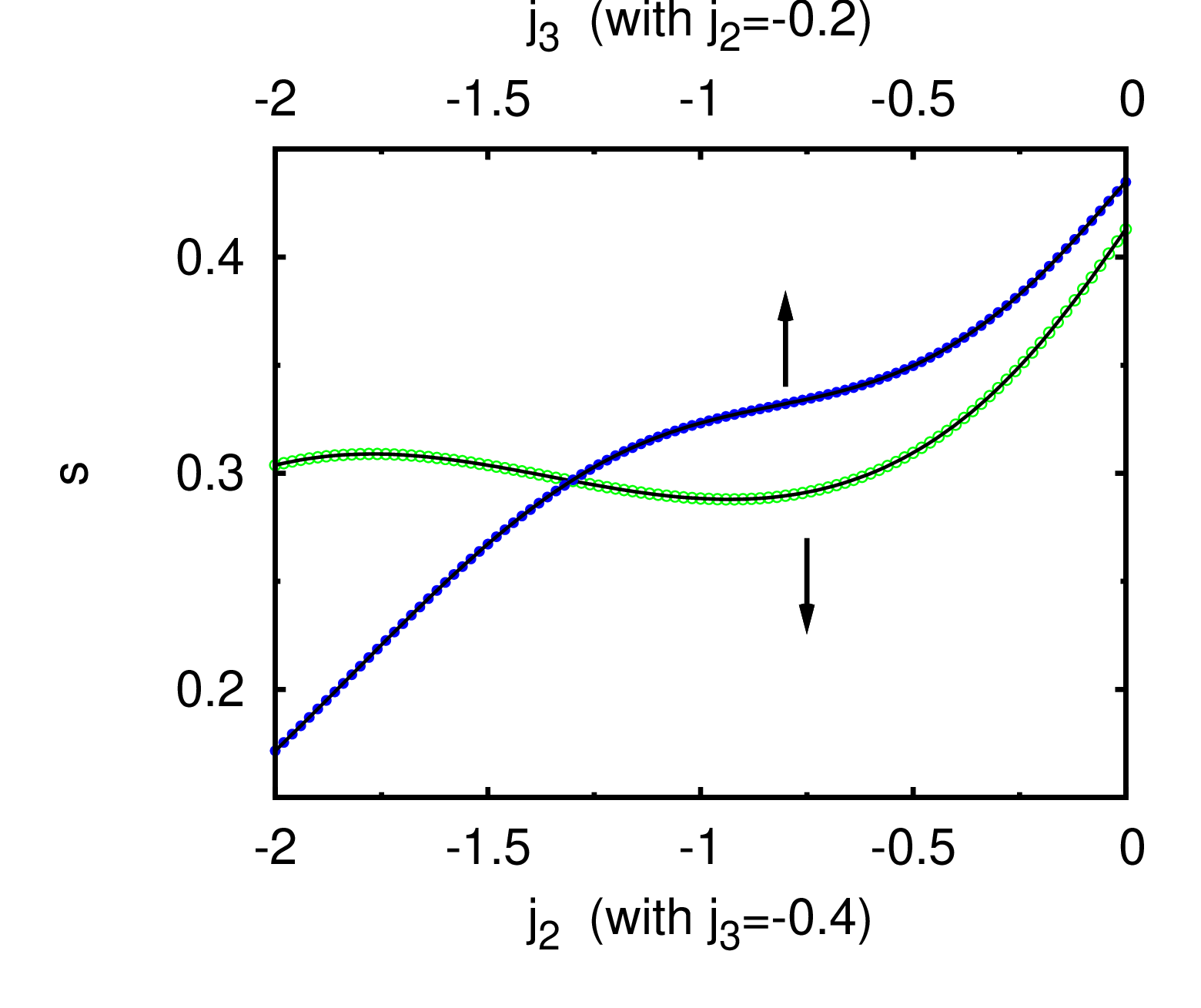}
\includegraphics[angle=270, width=.49\textwidth]{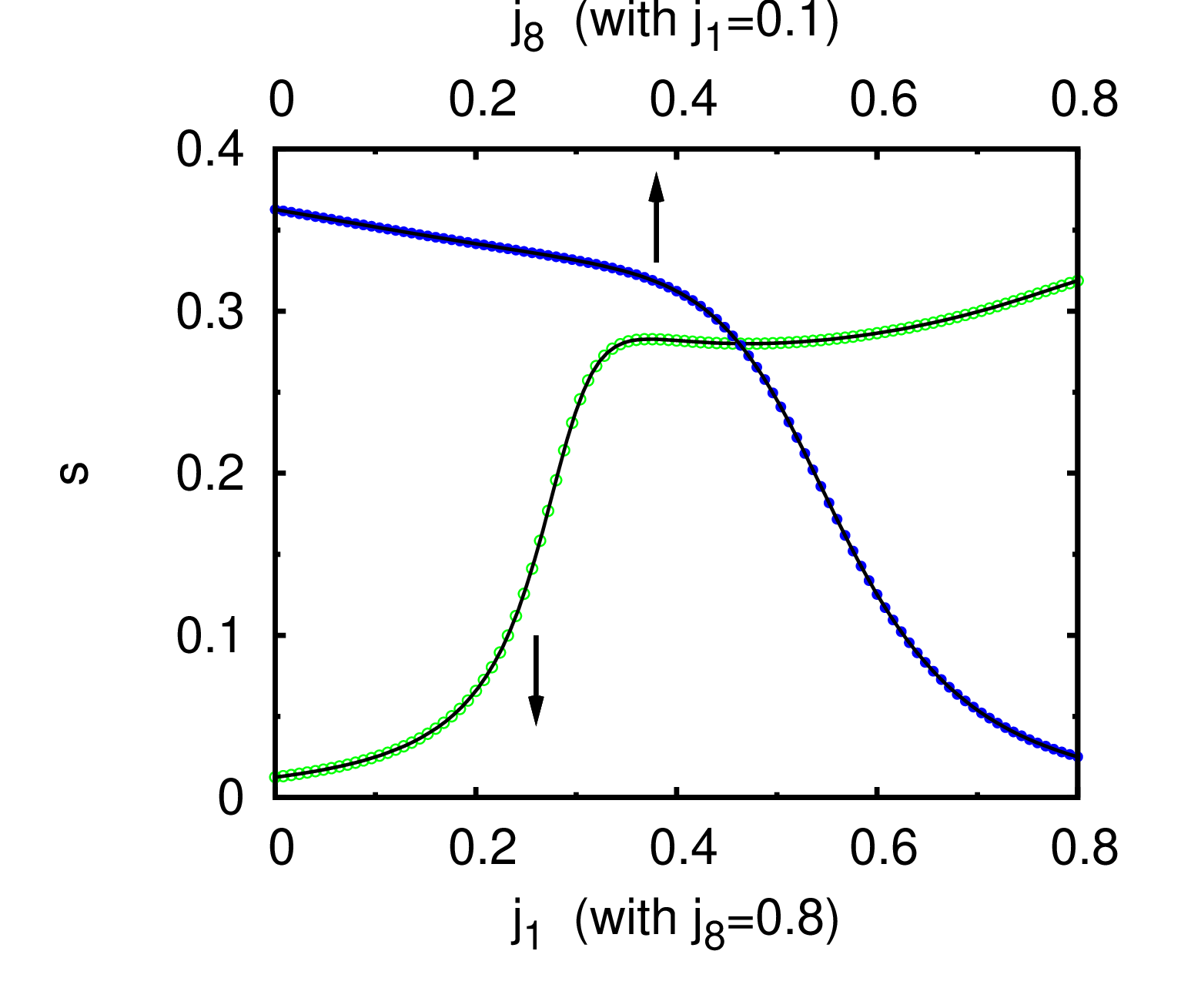}
\caption{Entropy for some values of the coupling constants: 
$j_1=-0.1, j_4=-0.8$ (left) for 
the  transfer matrix \eqref{TM2} and 
$j_2=0.2, j_3=-0.3, j_4=0.4, j_5=-0.5, j_6=0.6, j_7=-0.7$ (right) for 
the  transfer matrix \eqref{TM3}. The values of the remaining couplings are specified on the 
$x$-axes of the plots. The arrows point to the relevant $x$-axis. The 
continuous curves are obtained by numerical calculation, while the dots are calculated
by means of \eqref{entropy}.}
\label{TM4x48x8}
\end{center}
\end{figure}

Proceeding as before, we can 
obtain the entropy and correlation functions. 
The comparison between the resulting obtained findings 
and our inversion formula are 
shown in the left panel of figure \ref{TM4x48x8}.
The plot shows the numerically calculated entropy and the analytical 
one \eqref{entropy} 
where the numerical correlations are plugged in, for some values of the coupling constants.
The entropy in this case is given by the expression 
(the subscripts in the 
$g$'s just indicate to which coupling they are conjugated):
\begin{align}
&s(g_1,g_2,g_3,g_4)=-\frac{1+g_1-g_3-g_4}{4}\log\left(\frac{1+g_1-g_3-g_4}{8}\right)
-\frac{1+g_1-2g_2+g_3-g_4}{8}\log\left(\frac{1+g_1-2g_2+g_3-g_4}{8}\right)\nonumber\\
&-\frac{1-3g_1+2g_2+g_3-g_4}{8}\log\left(\frac{1-3g_1+2g_2+g_3-g_4}{8}\right)
-\frac{1-g_1-g_3+g_4}{4}\log\left(\frac{1-g_1-g_3+g_4}{8}\right)\nonumber\\
&-\frac{1-g_1-2g_2+g_3+g_4}{8}\log\left(\frac{1-g_1-2g_2+g_3+g_4}{8}\right)
-\frac{1+3g_1+2g_2+g_3+g_4}{8}\log\left(\frac{1+3g_1+2g_2+g_3+g_4}{8}\right)\nonumber\\
&+\frac{1-g_2}{2}\log\left(\frac{1-g_2}{4}\right)
+\frac{1-2g_1+g_2}{4}\log\left(\frac{1-2g_1+g_2}{4}\right)
+\frac{1+2g_1+g_2}{4}\log\left(\frac{1+2g_1+g_2}{4}\right)\nonumber.
 \end{align}

As a last case we consider a translation invariant chain with range 
$R=4$: identifying the states as  
$|\uparrow\uparrow\uparrow\rangle = (1,0,0,0,0,0,0,0)$,
$|\uparrow\uparrow\downarrow\rangle = (0,1,0,0,0,0,0,0)$,
$|\uparrow\downarrow\uparrow\rangle = (0,0,1,0,0,0,0,0)$,
$|\uparrow\downarrow\downarrow\rangle = (0,0,0,1,0,0,0,0)$,
$|\downarrow\uparrow\uparrow\rangle = (0,0,0,0,1,0,0,0)$,
$|\downarrow\uparrow\downarrow\rangle = (0,0,0,0,0,1,0,0)$,
$|\downarrow\downarrow\uparrow\rangle = (0,0,0,0,0,0,1,0)$,
$|\downarrow\downarrow\downarrow\rangle = (0,0,0,0,0,0,0,1)$
and the couplings as 
$j_{\{1\}}=j_1$, $j_{\{1,2\}}=j_2$, $j_{\{1,3\}}=j_3$, $j_{\{1,4\}}=j_4$, $j_{\{1,2,3\}}=j_5$, $j_{\{1,2,4\}}=j_6$, $j_{\{1,3,4\}}=j_7$, $j_{\{1,2,3,4\}}=j_8$, 
the $8$ by $8$ transfer matrix reads:
\begin{align}
\mathbf{T}&=\left(\begin{array}{cc}
\mathbf{T}_1&\mathbf{T}_2\\
\mathbf{T}_3&\mathbf{T}_4\\
\end{array}
\right)\label{TM3}
\end{align}
where the $4 \times 4$ matrices 
$\mathbf{T}_1,\cdots,\mathbf{T}_4$ are 
{\tiny
\begin{align}
\mathbf{T}_1&=\left(\begin{array}{cccc}
e^{3j_1+3j_2+3j_3+3j_4+3j_5+3j_6+3j_7+3j_8}&e^{3j_1+j_2+j_3+j_4-j_5-j_6-j_7-3j_8}&
e^{3j_1+3j_2+j_3+j_4+j_5+j_6-j_7-j_8}&e^{3j_1+j_2-j_3-j_4+j_5-3j_6-j_7+j_8}\\
e^{j_1+j_2+j_3+j_4+j_5+j_6+j_7+j_8}&e^{j_1-j_2-j_3+3j_4-3j_5+j_6+j_7-j_8}&
e^{j_1+j_2-j_3-j_4-j_5-j_6-3j_7-3j_8}&e^{j_1-j_2-3j_3+j_4-j_5-j_6+j_7+3j_8}\\
e^{j_1-j_2+j_3+j_4-j_5-j_6+j_7-j_8}&e^{j_1-3j_2+3j_3-j_4-j_5-j_6+j_7+j_8}&
e^{j_1-j_2-j_3+3j_4-3j_5+j_6+j_7-j_8}&e^{j_1-3j_2+j_3+j_4+j_5+j_6-3j_7+j_8}\\
e^{-j_1+j_2-j_3-j_4+j_5+j_6-j_7+j_8}&e^{-j_1-j_2+j_3+j_4+j_5-3j_6+3j_7-j_8}&
e^{-j_1+j_2-3j_3+j_4-j_5+3j_6-j_7+j_8}&e^{-j_1-j_2-j_3+3j_4+3j_5-j_6-j_7-j_8}
\end{array}
\right)\nonumber\\
\mathbf{T}_2&=\left(\begin{array}{cccc}
e^{3j_1+3j_2+3j_3+j_4+3j_5+j_6+j_7+j_8}&e^{3j_1+j_2+j_3-j_4-j_5+j_6-3j_7-j_8}&
e^{3j_1+3j_2+j_3-j_4+j_5-j_6+j_7+j_8}&e^{3j_1+j_2-j_3-3j_4+j_5-j_6+j_7-j_8}\\
e^{j_1+j_2+j_3-j_4+j_5-j_6-j_7-j_8}&e^{j_1-j_2-j_3+j_4-3j_5+3j_6-j_7+j_8}&
e^{j_1+j_2-j_3-3j_4-j_5-3j_6-j_7-j_8}&e^{j_1-j_2-3j_3-j_4-j_5+j_6+3j_7+j_8}\\
e^{j_1-j_2+j_3-j_4-j_5-3j_6-j_7-3j_8}&e^{j_1-3j_2+3j_3-3j_4-j_5+j_6-j_7+3j_8}&
e^{j_1-j_2-j_3+j_4-3j_5-j_6+3j_7+j_8}&e^{j_1-3j_2+j_3-j_4+j_5+3j_6-j_7-j_8}\\
e^{-j_1+j_2-j_3-3j_4+j_5-j_6-3j_7-j_8}&e^{-j_1-j_2+j_3-j_4+j_5-j_6+j_7+j_8}&
e^{-j_1+j_2-3j_3-j_4-j_5+j_6+j_7+3j_8}&e^{-j_1-j_2-j_3+j_4+3j_5+j_6+j_7-3j_8}
\end{array}
\right)\nonumber\\
\mathbf{T}_3&=\left(\begin{array}{cccc}
e^{j_1-j_2-j_3+j_4-3j_5-j_6-j_7-3j_8}&e^{j_1+j_2-3j_3-j_4+j_5-j_6-j_7+3j_8}&
e^{j_1-j_2+j_3-j_4-j_5+j_6-j_7+j_8}&e^{j_1+j_2-j_3-3j_4-j_5+j_6+3j_7-j_8}\\
e^{-j_1-3j_2+j_3-j_4-j_5-3j_6+j_7-j_8}&e^{-j_1-j_2-j_3+j_4+3j_5+j_6-3j_7+j_8}&
e^{-j_1-3j_2+3j_3-3j_4+j_5-j_6+j_7+3j_8}&e^{-j_1-j_2+j_3-j_4+j_5+3j_6+j_7-3j_8}\\
e^{-j_1-j_2-3j_3-j_4+j_5-j_6-3j_7+j_8}&e^{-j_1+j_2-j_3-3j_4+j_5+3j_6+j_7-j_8}&
e^{-j_1-j_2-j_3+j_4+3j_5-3j_6+j_7+j_8}&e^{-j_1+j_2+j_3-j_4-j_5+j_6+j_7-j_8}\\
e^{-3j_1+j_2-j_3-3j_4-j_5+j_6-j_7-j_8}&e^{-3j_1+3j_2+j_3-j_4-j_5+j_6-j_7+j_8}&
e^{-3j_1+j_2+j_3-j_4+j_5-j_6+3j_7-j_8}&e^{-3j_1+3j_2+3j_3+j_4-3j_5-j_6-j_7+j_8}
\end{array}
\right)\nonumber\\
\mathbf{T}_4&=\left(\begin{array}{cccc}
e^{j_1-j_2-j_3+3j_4-3j_5+j_6+j_7-j_8}&e^{j_1+j_2-3j_3+j_4+j_5-3j_6+j_7+j_8}&
e^{j_1-j_2+j_3+j_4-j_5+3j_6-3j_7-j_8}&e^{j_1+j_2-j_3-j_4-j_5-j_6+j_7+j_8}\\
e^{-j_1-3j_2+j_3+j_4-j_5-j_6+3j_7+j_8}&e^{-j_1-j_2-j_3+3j_4+3j_5-j_6-j_7-j_8}&
e^{-j_1-3j_2+3j_3-j_4+j_5+j_6-j_7+j_8}&e^{-j_1-j_2+j_3+j_4+j_5+j_6-j_7-j_8}\\
e^{-j_1-j_2-3j_3+j_4+j_5+j_6-j_7+3j_8}&e^{-j_1+j_2-j_3-j_4+j_5+j_6+3j_7-3j_8}&
e^{-j_1-j_2-j_3+3j_4+3j_5-j_6-j_7-j_8}&e^{-j_1+j_2+j_3+j_4-j_5-j_6-j_7+j_8}\\
e^{-3j_1+j_2-j_3-j_4-j_5+3j_6+j_7+j_8}&e^{-3j_1+3j_2+j_3+j_4-j_5-j_6+j_7-j_8}&
e^{-3j_1+j_2+j_3+j_4+j_5+j_6+j_7-3j_8}&e^{-3j_1+3j_2+3j_3+3j_4-3j_5-3j_6-3j_7+3j_8}
\end{array}
\right)\nonumber
\end{align}
}
Results are shown in figure \ref{TM4x48x8} 
(we do not write down the entropy for this case).
By inspecting figure \ref{TM4x48x8} we find that the 
agreement between the two approaches is complete. We do not report here 
the other checks we performed for higher values of $R$.

\end{document}